\documentclass[aps,twocolumn,showpacs,floatfix,longbibliography]{revtex4-1}

\usepackage{txfonts}
\usepackage{microtype}
\usepackage{ulem}
\usepackage{epsfig}
\usepackage{amssymb}
\usepackage{bm}
\usepackage{graphicx}
\usepackage{color}
\usepackage[outdir=./fig/CMT/]{epstopdf}

\newcommand{\dd}{\mathrm{d}} 
\newcommand*{\sign}{\mathop{\mathrm{sign}}\nolimits}

\begin{document}
\author{Ji\v{r}\'{i} Dan\v{e}k}\email{danek@mpi-hd.mpg.de}
\author{Karen Z. Hatsagortsyan}\email{k.hatsagortsyan@mpi-hd.mpg.de}
\author{Christoph H. Keitel}
\affiliation{Max-Planck-Institut f\"ur Kernphysik, Saupfercheckweg 1, 69117 Heidelberg, Germany}

\bibliographystyle{apsrev4-1}

\title{Analytical approach to Coulomb focusing in strong field ionization}

\date{\today}

\begin{abstract}

The role of the Coulomb potential of the atomic core for creation of caustics in the photoelectron momentum distribution for tunneling ionization in a linearly polarized strong laser field, usually termed as Coulomb focusing, is investigated within classical theory beyond the dipole approximation. Coulomb focusing is addressed by analytical calculation of Coulomb momentum transfer to the tunneled electron due to rescatterings, 
while  applying perturbation theory and classifying the recollisions either as fast or as slow. The accuracy of the obtained analytical formulas for the Coulomb momentum transfer is investigated by analytical derivation of asymptotic photoelectron momentum distribution and its comparison to  exact numerical calculations. 
With the help of the applied analytical treatment, we analyze the origin of the counterintuitive energy-dependent bend of the Coulomb focusing cusp in the photoelectron momentum distribution in a linearly polarized laser field in the non-dipole regime, and its scaling with the field parameters. 
The importance of high-order recollisions is also investigated and they are shown to be responsible for a decrease of the bend of the cusp at very low energies in this regime. In general, the
momentum transfer caused by high-order rescattering events, while being a perturbation with respect to the instantaneous electron momentum at the rescattering, is shown to induce a significant non-perturbative contribution to the total momentum transfer.


\end{abstract}

\pacs{32.80.Rm,34.80.Bm}

\maketitle

\section{Introduction} \label{sec:dipole:01:introduction}

Since the seminal works of Perelomov, Popov, Terent'tev (PPT) \cite{Perelomov_1966b,Perelomov_1967a,Popov_1967,Popov_2004u,Popov_2005,Popruzhenko_2014r} it was known that the ionization rate of the atom in a strong laser field can be significantly disturbed by the Coulomb field of the atomic core. Later it has been realized that the Coulomb field of the atomic core imprints specific signatures on the momentum distribution of photoelectrons \cite{Brabec_1996}, which arise during electron excursion in the laser field after the release (tunneling) from the bound state. The Coulomb field effect on the electron dynamics is conspicuous, first of all,  near the tunnel exit \cite{Goreslavski_2004}, and further during rescatterings \cite{Corkum_1993}. While the first effect exists at any polarization of the laser field, the rescattering is mostly efficient in the case of linear polarization. Although even in a laser field of elliptical plarization rescattering and consequent Coulomb effects can take place \cite{Shvetsov-Shilovski_2008,Wang_2009,Wang_2010,Mauger_2010a,Chengpu_2012,Mauger_2013,Dimitrovski_2014,Dimitrovski_2015}.

Hard rescatterings with a small impact parameter induce well-known processes of above-threshold ionization \cite{Becker_2002}, high-order harmonic generation \cite{Agostini_2004, KOHLER2012159}, and  nonsequential double ionization \cite{Becker_2012}. In contrast to that, due to multiple forward scattering of ionized electrons by the atomic core at large impact parameters during oscillation in the laser field, the electrons with large transverse momentum at the ionization tunnel exit finally appear with low transverse momentum. Accordingly, the large initial transverse momentum space at the tunnel exit is squeezed into the asymptotic small one, i.e., the Coulomb field focuses electrons in the momentum space along the laser polarization direction which is termed as Coulomb focusing (CF)
\cite{Brabec_1996,Yudin_2001a,Comtois_2005}. 
In early experiments the traces of CF was observed  as cusps and humps in the photoelectron momentum distribution (PMD) \cite{Moshammer_2003,Rudenko_2004,Dimitriou_2004,Rudenko_2005}. Recently, due to advancements of mid-infrared laser technique \cite{Wolter_2015x}, the interest to CF has been significantly increased with observation of rich structures in PMD near the ionization threshold in long wavelength laser fields, the, so-called, low-energy structures (LES) \cite{Blaga_2009,Catoire_2009,Quan_2009}, very-low-energy structures  \cite{Wu_2012b,Wolter_2014}, and zero-energy structures  \cite{Dura_2013,Pullen_2014,Xia_2015,Zhang_2016,Diesen_2016,Williams_2017}. The origin of LES has been traced to multiple  forward scattering  by the Coulomb field, which induces transverse and longitudinal bunching of the electron momentum space \cite{Faisal_2009,Liu_2010,Yan_2010,Guo_2010,Liu_2011,Kastner_2012,Lemell_2012,Guo_2013}.
LES are well resolved in mid-infrared laser fields, when the Keldysh parameter is small, the interaction is essentially in the tunneling regime, and when classical features of the three-step
model \cite{Corkum_1993} are evident.

For nonperturbative quantum descriptions of Coulomb field effects, different modifications of
the strong field approximation (SFA) \cite{Keldysh_1965,Faisal_1973,Reiss_1980} have been developed
\cite{Popruzhenko_2008a,Popruzhenko_2008b,Torlina_2012,Torlina_2012b,Kaushal_2013,Klaiber_2013a,Klaiber_2013b,Lai_2015}.  The Coulomb-corrected SFA (CCSFA) of \cite{Popruzhenko_2008a,Popruzhenko_2008b}, which employs the quasiclassical electron wave function in the continuum in laser and Coulomb fields, has been successfully applied for the explanation of LES \cite{Yan_2010}. A similar but more systematic R-matrix theory (ARM) \cite{Torlina_2012,Torlina_2012b,Kaushal_2013} also has been extended to treat recollisions \cite{Pisanty_2016,Keil_2016}.  It appeared  that the perturbative SFA is also able to account for LES \cite{Milosevic_2013,Milosevic_2014a,Becker_2014,Moller_2014,Becker_2015,Kelvich_2016}, when appropriate trajectories with soft recollisions \cite{Kastner_2012} are included. However, this description is only qualitative, because for a correct quantitative description the effect of multiple recollisions should be taken into account.

In mid-infrared laser fields, the electron dynamics after tunneling is mainly classical, because the characteristic energies of the process, namely the ionization and ponderomotive potentials, greatly exceed the photon energy in this regime. Therefore the classical trajectory Monte Carlo (CTMC) method \cite{Leopold_1979,Hu_1997,Cohen_2001} has been successful in explaining LES features, see e.g., \cite{Quan_2009,Liu_2010,Lemell_2012,Hickstein_2012,Liu_2014a}.
Although both CCSFA and CTMC successfully predict the existence of LES,  they deliver only little insight into the underlying physics as they both employ classical trajectories via numerical calculations which hide the physical picture of the transformation of the electron's initial momentum space at the tunnel exit into the asymptotic one at the detector.

CF arises due to the long range Coulomb interaction between the tunnelled electron and its parent ion. This interaction is conspicuous  at rescattering points when the tunneled electron revisits the atomic core during its excursion driven by the laser field. Usually the momentum transfer during high-order rescattering events is decreasing with its order. However, the decrease is not monotonous and the accurate description requires accounting also for high-order rescatterings \cite{Liu_2011,Hickstein_2012}.

Moreover, recent experiments \cite{Ludwig_2014,Maurer_2017} have shown that CF is significantly modified in the non-dipole regime. The breakdown of the dipole approximation was firstly observed in the case of linear polarization of the laser field \cite{Ludwig_2014}, as a counterintuitive shift of the PMD peak opposite to the laser propagation direction which was attributed to the interaction of the tunneled electron with the Coulomb field of the parent ion. Further numerical calculations of time-dependent Schr\"odinger equation have shown that the PMD shift with respect to the dipole approximation case is not uniform  but momentum dependent \cite{Bandrauk_2015}. The same conclusion has been drawn from the classical \cite{Tao_2017}, and CCSFA calculations \cite{Keil_2017}. However, the intuitive explanation of the non-dipole features of PMD  is still missing.

In this paper, we develop a classical analytical theory for the description of CF with respect to the underlying momentum transfer due to the Coulomb interaction. We derive analytical formulas for the Coulomb momentum transfer to the recolliding electron at multiple recollisions, while classifying the recollisions as either fast or slow recollision. 
We include non-dipole effects, accounting for the laser magnetic field induced drift of the  ionized electron along the laser propagation direction during the excursion in the laser field.
The Coulomb field of the atomic core is treated as a perturbation to the laser driven trajectory near the recollision point. The scaling of the Coulomb momentum transfer at the recollision (R-CMT) with respect to the rescattering parameters (momentum and impact parameter) as well as with respect to the laser intensity and wavelength is derived. Special attention is devoted to the contribution of high-order rescattering events, and to the derivation of the effective number of rescatterings. Furthermore, we provide  high-order corrections to the known formula for the initial Coulomb momentum transfer (I-CMT) which 
is necessary for keeping the overall precision of our  model.

Once seizing the Coulomb interaction through I-CMT and R-CMT, 
we estimate the total Coulomb momentum transfer and derive the final PMD. Two 
methods are applied: fully perturbative, and step-by-step method. While in the first method the Coulomb effect is assumed to be a perturbation with respect to the laser driven global electron trajectory, in the second method the trajectory is adjusted after each recollision.

Finally, we employ our framework for investigation of the counterintuitive energy-dependent bend of the cusp in the PMD, while revealing a fine interplay of the non-dipole and Coulomb field effects.
We find a direct relationship of multiple recollisions and the fine structure of the cusp.


The structure of the paper is the following. In Sec. \ref{sec:dipole:02:introducing_model} the model of CF is introduced. In Sec. \ref{sec:dipole:03:distinctive_traj} the characteristic trajectories 
at CF are discussed. 
The transverse and longitudinal R-CMT are calculated in all generality in Sec. \ref{sec:dipole:04:solving_first_order},  and in the Sec. \ref{sec:dipole:05:intial_mom_transfer}, we derive the  revised formula for I-CMT. 
The total Coulomb momentum transfer and the final PMD with two methods are derived in Sec. \ref{sec:dipole:07:total} and 
the accuracy of our methods is investigated
in the case of the dipole approximation. 
The 
The non-dipole effects in CF are investigated in Sec.\ref{sec:non-dipole:07:non-dipole}. 
The conclusion is given in Sec. \ref{sec:dipole:08:discussion}.

\section{The model}
\label{sec:dipole:02:introducing_model}

We consider the tunneling ionization regime of an atom in a strong laser field  when the Keldysh parameter is small \cite{Keldysh_1965}, i.e. , $\gamma\equiv\sqrt{I_p/2U_p}\ll 1$, with the ionization potential $I_p$ and the ponderomotive potential $U_p=E_0^2/4\omega^2$. Atomic units are used throughout, unless mentioned otherwise. The laser field is linearly polarized 
\begin{eqnarray}
\mathbf{E}(u) &=& E_0\mathbf{e}\cos{u},\\
\mathbf{B}(u)&=&\textbf{n}\times \textbf{E}(u),\nonumber
\end{eqnarray}
where $u=\omega (t-z/c)$ is the laser phase,  $\mathbf{B}(u)$ is the laser magnetic field, $E_0$ and $\omega$ are the amplitude and the angular frequency of the laser field, respectively, $c$ is the speed of light, $\mathbf{e}=(1,0,0)$ and $\mathbf{n}=(0,0,1)$ are the unit vectors along the laser polarization and propagation directions, respectively. 
 We assume that the electron has tunneled out from the atomic bound state which is described by the PPT ionization rate. The latter provides probability for the initial transverse momentum distribution \cite{Popov_2004u}
\begin{eqnarray}
w(p_\perp)\propto \exp\left(-\frac{p_\perp^2}{\Delta_\perp^2}\right),
\end{eqnarray}
where $\Delta_\perp=E_0^{1/2}/ (2I_p)^{1/4}$. The initial longitudinal momentum 
with respect to the laser field direction at the   ionization moment $t_i$ is assumed to be vanishing, and the initial coordinate is at the tunnel exit.

We consider the non-dipole regime of interaction and will keep for the  solution of the equations of motion the leading terms with respect to $1/c$, which describe the laser magnetic field induced drift of the electron in  the laser propagation direction. The physical condition of the applied $1/c$-expansion is the smallness of the laser induced drift distance during the laser period $d\sim \lambda\xi^2 /2$ \cite{RMP_2012} with respect to the recollision impact parameter $\rho \sim 2\pi p_\bot/ \omega$: $d\ll \rho$, where $\xi=E_0/(c\omega)$ is the invariant laser field parameter, $\lambda$ is the laser wavelength,  $p_\bot=\sqrt{p_y^2+p_z^2}$ is the electron transverse momentum, $p_\bot\sim 2\Delta_\bot=2\kappa \sqrt{E_0/E_a}$, $\kappa=\sqrt{2I_p}$ is the atomic momentum, $I_p$ is the ionization potential, and $E_a=\kappa^3$ is the atomic field. Note that the introduced small parameter $\epsilon\equiv d/\rho$, in fact, is directly related to the Lorentz deflection parameter \cite{Walker_2006,Klaiber_2017}:
\begin{eqnarray}
\Gamma_R=\epsilon^2=\frac{\kappa c\xi^3}{16\omega}.\label{d-rho}
\end{eqnarray}

The magnetically induced drift changes the impact parameter of recollisions and in this way modifies CF. 
However, we  stress that during the brief recollision time $\delta t$  the effect of the magnetically induced drift is negligible, because the change of the impact parameter due the drift during the recollision time, which can be estimated as $\delta \rho \sim(\lambda \xi^2) (
\omega \delta t)$, is much smaller than the impact parameter itself. In fact, we estimate the recollision time as $\delta t\sim \rho/v_{\|}$, with the electron longitudinal velocity at the recollision $v_{\|}\sim E_0/\omega$, and the ratio $\delta \rho/\rho\sim \epsilon\gamma \sqrt{E_0/E_a}$. We consider the tunneling regime when the Keldysh parameter is small $\gamma\ll 1$, and the field is small to hinder the over-the-barrier ionization, i.e.,  $E_0/E_a\ll 1$. The latter means that $\delta \rho/\rho$, the change of the impact parameter due the drift during the recollision time, has  an additional smallness  in addition to the small parameter $\epsilon$ and, consequently, can be neglected in our discussion.

Our aim is to find an analytical expression for the Coulomb momentum transfer. We assume that the Coulomb field effect is not negligible only near recollision points and near the tunnel exit, where it is treated as a perturbation with respect to the laser field. The latter assumptions are valid if, firstly, the Coulomb force is smaller with respect to the laser field at the recollision point and at the tunnel exit: $Z/r_r^2\, , Z/x_e^2\ll E_0$, the charge of the atomic core $Z$, the recollision and the tunnel exit coordinates $r_r\sim \Delta_\bot /\omega$, and  $x_e\sim I_p/E_0$, respectively, and secondly, if the quiver amplitude of the electron in the laser field greatly exceeds the recollision and the tunnel exit coordinates $E_0/\omega^2\gg r_r\, , x_e$. The first pair of these conditions reads
\begin{eqnarray}
\frac{Z}{\kappa}\gamma^2 & \ll & 1,\\
\frac{Z}{\kappa}\frac{E_0}{E_a}&\ll & 1,
\end{eqnarray}
and the second pair gives
\begin{eqnarray}
 \gamma\sqrt{\frac{E_0}{E_a}}&\ll& 1,\\
 \gamma^2 & \ll & 1.
\end{eqnarray}
These conditions are well fulfilled in the tunneling regime.

The tunnelled electron dynamics in the continuum after tunneling is governed by Newton equations 
\begin{eqnarray}
\frac{d\mathbf{p}}{dt} = -\mathbf{E}-\frac{\textbf{v}}{c}\times\mathbf{B} -\frac{Z\mathbf{r}}{{r}^3},\label{newton1}
\end{eqnarray}
where  $\textbf{v}$ is the electron velocity. The Coulomb field of the atomic core will be treated by perturbation theory during the recollision  and we expand the momentum and coordinate as
\begin{eqnarray}
\mathbf{p}&=&\mathbf{p}_0+\mathbf{p}_1+\ldots \,\,,\nonumber\\
\mathbf{r}&=&\mathbf{r}_0+\mathbf{r}_1+\ldots \,\,.
\end{eqnarray}
The unperturbed trajectory $\mathbf{r}_{0}(u)$ is determined by the laser field
\begin{equation}
\frac{d\mathbf{p}_{0}}{dt} = -\mathbf{E}\left(1-\frac{\textbf{n}\cdot\textbf{v}_{0}}{c}  \right)-\textbf{n}\frac{\textbf{v}_{0}\cdot\textbf{E}}{c},\label{eq:01:perturbative_newton}
\end{equation}
and momentum transfer due to the Coulomb field at the recollision is described by the trajectory in the first order of perturbation
\begin{eqnarray}
 \frac{d\mathbf{p}_{1}}{dt}  &=& -\frac{Z\mathbf{r}_{0}}{ r_{0}^3},
\end{eqnarray}
with $r_{0}=\left|\mathbf{r}_{0}\right|$. Taking into account that $du/dt=\omega(t-v_z/c)$, and the integral of motion in a plane laser field $\Lambda_0\equiv \varepsilon_0(u) -cp_{0z}(u)=\rm const$, with the electron energy $\varepsilon_0$,   Eq. ~(\ref{eq:01:perturbative_newton}) are integrated, providing the laser driven momentum evolution
\begin{eqnarray}
p_{0x}(u) 	& = & p_{x r}  + \left[A_x(u)-A_x(u_r)\right] ,\nonumber\\
p_{0y}(u) 	& = & p_{y r},	\label{eq:01:zero_order:mometum}\\
p_{0z}(u) & = & p_{z r}+p_{zd}(u,u_r),
\nonumber
\end{eqnarray}
with the laser vector-potential $A_x(u)=-(E_0/\omega)\sin u$. The initial conditions are defined at the recollision point with the recollision phase $u_r$, and the recollision momentum $\textbf{p}_r=(p_{x r},p_{y r},p_{z r})$, 
aiming at application of  the solution near the recollision point. Here, the drift momentum induced by the laser magnetic field is
\begin{eqnarray}
p_{zd}(u,u_r) \equiv \frac{p_{x r}}{c}\left[A_x(u)-A_x(u_r)\right]+\frac{1}{2c}\left[A_x(u)-A_x(u_r)\right]^2,\nonumber \\
\label{eq:pz_drift_recollpoint}
\end{eqnarray}
where  the 
integral of motion is approximated $\Lambda_0\approx c^2$, to keep the leading term in $1/c$ expansion. The unperturbed electron trajectory near the recollision point is
\begin{eqnarray}
x_0(u) 	& = &  \frac{E_0}{\omega^2}\left[
\cos{u} - \cos{u_r} + (u - u_r) \sin{u_r}
\right] +\frac{p_{xr}}{\omega}(u - u_r)+ x_r.\nonumber\\
y_0(u) 	& = & \frac{p_{y r}}{\omega}(u-u_r)+y_{r},\label{eq:01:02:zero_order:z}	\\
z_0(u) 	& = & \frac{p_{z r}}{\omega}(u-u_r)+z_d(u)+z_{r},\nonumber 
\end{eqnarray}
with the recollision coordinate $\textbf{r}_r=(x_r,y_r,z_r)$, and the laser magnetically induced drift coordinate
\begin{eqnarray}
z_d(u)=\int_{u_r}^up_{zd}(u',u_r) \dd u'.\label{zd}
\end{eqnarray}

Once  the zero-order equations are solved, the momentum transfer due to the Coulomb field at the  recollision can be derived as the first-order correction
\begin{eqnarray}
  \textbf{p}_{1 }(u) 		=
   -\frac{Z}{\omega}\int\limits_{u_r-\delta}^{u_r+\delta} \frac{\textbf{r}_0(u^{\prime})}{r_0^3(u^{\prime})}\dd u^{\prime},	\label{eq:01:03:first_order:momentum_trans}
   \end{eqnarray}
where $\textbf{r}_0(u) =\left(x_0(u),y_0(u),z_0(u)\right)$. The value of the parameter $\delta$  is coupled to the properties of the recollision and will be discussed in the next section.

In the discussion above, the Coulomb field effect is accounted  for only near recollision points and near the tunnel exit, where it is treated as a perturbation with respect to the laser field. Still Eq.~(\ref{eq:01:03:first_order:momentum_trans}) for the Coulomb momentum transfer (CMT) includes nonperturbative Coulomb effects via  dependence on the recollision parameters, i.e., the electron momentum $\textbf{p}_r$ and coordinate $\textbf{r}_r$ at the recollision point. 
In fact, the multiple recollisions preceding the currently discussed  one can have significant contribution to the considered $\textbf{p}_r$ and $\textbf{r}_r$, although the Coulomb field is perturbation at every single one of them. Nevertheless, we will concentrate only on the single recollision for now and return to the role of multiple recollisions in Sec. \ref{sec:dipole:07:total} where we address the role of multiple recollisions for the total momentum transfer.

One may apply also a less accurate description assuming that the Coulomb field is perturbation globally, i.e, at any moment the difference between the exact and laser driven trajectories is a perturbation. In this description the unperturbed electron trajectory is given by  Eqs.~(\ref{eq:01:zero_order:mometum})-(\ref{zd}), replacing the recollision phase $u_r$ by the ionization phase $u_i$, and the recollision coordinate $\textbf{r}_{r}$,  and momentum $\textbf{p}_{r}$  by the coordinate and momentum at the tunnel exit:  $\textbf{r}_i=\left(-I_p/E(u_i),0,0\right)$, and $\textbf{p}_{i}=\left(0,p_{yi},p_{zi}\right)$.

Although generally we do not apply dipole approximation, we will check the accuracy of our method in Sec.~\ref{sec:dipole:07:total} in the dipole approximation case, which will be specially indicated.

\section{Classification of recollisions}\label{sec:dipole:03:distinctive_traj}

For estimation of R-CMT we firstly classify recolliding trajectories. There are two characteristic recolliding trajectories: 1) when the electron longitudinal velocity is vanishing at the recollision point, $x_r=0$ and $p_{xr}=0$, and 2) when the electron has the highest velocity at the recollision point, $x_r=0$, $|\textbf{p}_r|\neq 0$ and $\textbf{E}(u_r)=0$. We will call the above mentioned two types of  recollisions as a slow recollision (SR), and a fast recollision (FR), respectively. 
The first type of recollision corresponds to the peak of the momentum transfer in dependence on the initial ionization phase, and the second type corresponds to the plateau of the momentum transfer, as discussed in \cite{Liu_2011}.
Let us remark that the SR is also known in the literature as ``soft recollision'' \cite{Kastner_2012}.

\begin{figure}[tb]
	\begin{center}
	\includegraphics[width = 0.8\linewidth]{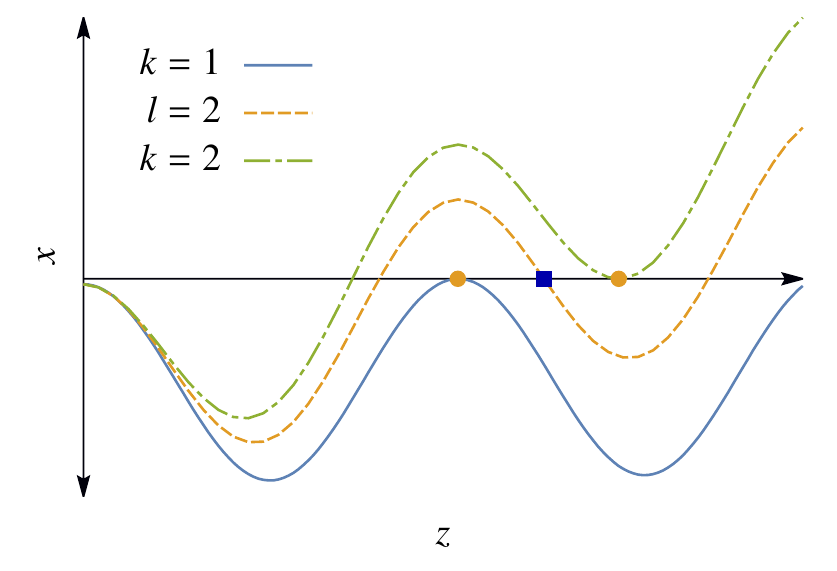}
	\end{center}
  		    \caption{Green (dot-dashed) and blue (dashed) trajectories correspond to SR with $k=1$ and $k=2$ (yellow circles), respectively. Yellow trajectory corresponds to FR with $l=2$ (blue square).}
    \label{fig:01:kinds_of_trajectories}
\end{figure}

In the case of the global perturbation for the Coulomb field, one uses the zero-order solutions  Eqs. (\ref{eq:01:02:zero_order:z}), and finds the recollision phases.
The SR phase $u_s$ is determined by the conditions
\begin{eqnarray}
x_0( u_{s}) & = & 0, \label{eq:02:peak_def}\\
x^{\prime}_0( u_{s}) & = & 0, \\
x^{\prime\prime}_0( u_{s}) & \neq & 0,
\end{eqnarray}
where prime denotes derivative with respect to the phase $u$, while the FR phase $u_f$ fulfills:  
\begin{eqnarray}
 x_0( u_{f}) & = & 0, \label{eq:03:plateau_def1}\\
x^{\prime}_0( u_{f}) & \neq & 0,\\
x^{\prime\prime}_0( u_{f}) & = & 0.   \label{eq:03:plateau_def}
\end{eqnarray}
 The condition $x_{0}^{\prime}(u)=0 $ for the $k^{\text{th}}$-SR leads to
\begin{equation}
u_{s}^{(k)} = (-1)^{k+1} u_i + \pi(k+1),\,\,\,\,k \in \mathbb{N}
\label{eq:04:00:peak_ur}
\end{equation} 
Odd values of $k$ correspond to trajectories approaching the ion from below, whereas even values to trajectories returning to the ion from above, see Fig.~\ref{fig:01:kinds_of_trajectories}. 
The ionization phase $u_{i,s}^{(k)}$ leading to the $k^{\text{th}}$-SR  is derived from Eq.~(\ref{eq:02:peak_def}):
\begin{eqnarray}
&&\frac{E_0}{\omega^2} \sin{u_i}\left[\pi(k+1)\right]  + x_i = 0, \,\,\  \text{for } k\text{ odd},\label{eq:03:01:peak:condition_for_ui}\\
&&\frac{E_0}{\omega^2}\left\{-2\cos{u_i} + \left[-2u_i+\pi(k+1)\right]\sin{u_i} \right\} + x_i =0,  \text{ for } k\text{ even}.\nonumber \\
\label{eq:04:01:peak:condition_for_ui}
\end{eqnarray}
Note that the tunnel exit $x_i(u_i)$ also depends on the ionization phase via $E(u_i)$. 
Taking into account that the ionization is most probable to take place near the maximum of the field $u_i\approx 0$, we  expand Eqs. (\ref{eq:03:01:peak:condition_for_ui}) and (\ref{eq:04:01:peak:condition_for_ui}) over $u_i\ll 1$, which leads us to the solution
\begin{eqnarray}
u_{i,s}^{(k)} &\approx& \frac{1+\left(-1\right)^k}{\pi
\left(k+1\right)} - \frac{x_i \omega^2}{\pi\left(k+1\right)E_0}\nonumber\\
&\approx& \frac{1+\left(-1\right)^k}{\pi
\left(k+1\right)},
\label{eq:04:peak_ui}
\end{eqnarray}
where  in the second step, it is taken into account that the term $x_i\omega^2/E_0= I_p\gamma^2\ll 1$ is small in the deep tunnelling regime  and can be neglected. 

The phase of FR is derived from  the condition $x^{\prime\prime}_0( u_{f})=0$:
\begin{equation}
u_{f}^{(l)} = \pi/2 + \pi l, \,\,\,\,\, l \in \mathbb{N}
\label{eq:04:00:plateau_ur}
\end{equation} 
A direct substitution into the  condition in Eq.~(\ref{eq:03:plateau_def1}) yields
\begin{equation}
\frac{E_0}{\omega}\left[
-\cos{u_i} + \sin{u_i}\left(\frac{2l+1}{2}\pi - u_i\right)
\right] + x_i = 0,
\label{eq:04:01:plateau:condition_for_ui}
\end{equation}
which can be again solved in the approximation of small ionization phase, providing the initial ionization phase $u_{i,f}^{(l)}$ leading to the $l^{\text{th}}$-FR 
\begin{eqnarray}
u_{i,f}^{(l)} \approx \frac{2}{\pi(2l+1)}\left(1-\frac{x_i\omega^2}{E_0}\right) 
\approx \frac{2}{\pi(2l+1)},
\label{eq:04:plateau_ui}
\end{eqnarray}
where the small term in the brackets $x_i\omega^2/E_0= I_p\gamma^2 $ is again neglected.

\section{Momentum transfer during recollisions}
\label{sec:dipole:04:solving_first_order}

The  momentum transfer to the recolliding electron due to the Coulomb field of the atomic core  at the $k^{\text{th}}$ recollision event is given by Eq.~(\ref{eq:01:03:first_order:momentum_trans}).
As the main contribution to the integrals comes near the recollision points, we expand the trajectory near the recollision phase where R-CMT takes place. The trajectory of the electron of Eq.~(\ref{eq:01:02:zero_order:z}) at the recollision point  $(x_r,y_r,z_r)$, with the recollision momentum $(p_{x r}, p_{y r},p_{z r})$
we approximate near the recollision phase $u_r$ as an expansion in  $\sigma=u-u_r$, up to the $\sigma^2$-order:
\begin{eqnarray}
x_0(u) &\approx& x_r  + \frac{p_{x r}}{\omega}\sigma -\frac{E(u_r)}{2\omega^2}\sigma^2\nonumber\\
 y_0(u) &\approx & y_r  + \frac{p_{y r}}{\omega} \sigma ,\\
 z_0(u) &\approx & z_r  + \frac{p_{z r}}{\omega}\sigma -\frac{p_{xr}E(u_r)}{2c\omega^2}\sigma^2\nonumber,
\end{eqnarray}
because $z_d(u_r)=0$, and $z'_d(u_r)=p_{zd}(u_r)=0$. 

\subsection{Slow recollisions}\label{sec:SR}

In the case of SR  $p_{x r}=0$, consequently, the trajectory in the leading order is
\begin{eqnarray}
x_0(u) \approx x_r- \frac{E(u_r)}{2\omega^2}\sigma^2, \,\,\,\,
y_0(u) \approx  y_r ,\,\,\,\,\,\,\,\,z_0(u) \approx  z_r.
\label{eq:SR_leading_order_x_z}
\end{eqnarray}
In the latter we have neglected $p_\bot\sigma/\omega$ terms with respect to the recollision coordinate $\rho \sim p_\bot /\omega$, because the effective value of $\sigma$,  derived from the condition $E_0\sigma^2/\omega^2 \sim \rho\sim p_\bot /\omega$ is $\sigma\sim \sqrt{p_\bot\omega/E_0}\sim \sqrt{\gamma \sqrt{E_0/E_a}}\ll 1$, 
and the transversal motion near the recollision point can be neglected for SR. 
In general, the rescattering parameter $x_r$ neither has to vanish at rescattering recollision nor has to be small. This assumption leads to generalization of the SR recollision for a larger class of recollisions with vanishing velocity. On the other hand, the extension to non-vanishing $x_r$ requires a bit more caution and will be discussed in further details in Sec. \ref{sec:dipole:07:total}.
\begin{figure}[b]
	\centering
  		  		\includegraphics[width=0.7\linewidth]{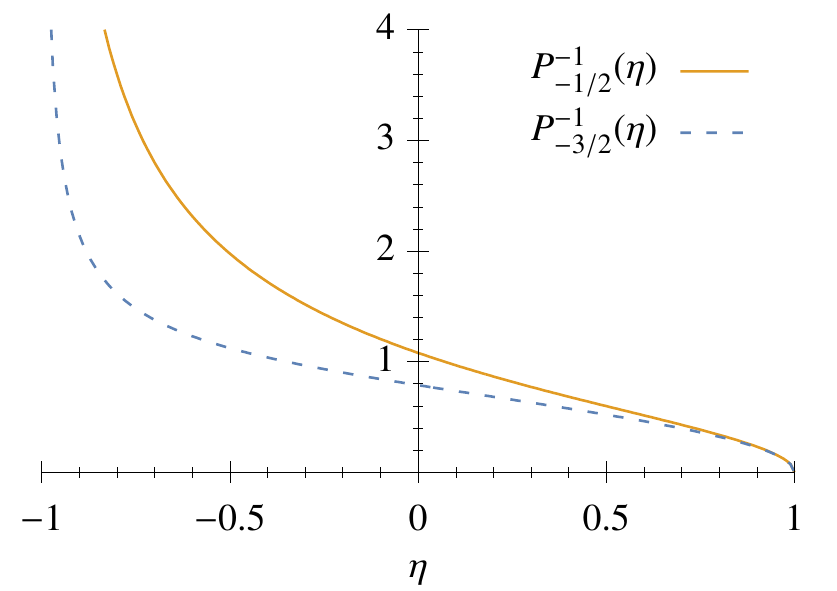}
\caption{The Legendre functions within the valid range given by Eqs.~(\ref{eq:14:corr_peak_preciser_generealized_long1})-(\ref{eq:14:corr_peak_preciser_generealized_trans1}). Distinct values are $P_{-3/2}^{-1}(0)\approx 0.787$ and $P_{-1/2}^{-1}(0)\approx 1.08$. Both functions diverge at $\eta\rightarrow -1$, which is out of the region of our interest since physically relevant cases correspond to $|\eta|<1/\sqrt{2}$. }
\label{fig:08:Legendre}
\end{figure}

From Eqs.~(\ref{eq:01:03:first_order:momentum_trans}), we calculate R-CMT for SR along the  the trajectory approximated by Eqs.~(\ref{eq:SR_leading_order_x_z}) while extending the integration limits to infinity ($\delta \rightarrow \infty$). The later is well justified since large values of $\sigma$ correspond to large deviation of the phase from the recollision point giving negligible contribution to the integration which now yields
\begin{eqnarray}
p_{1 x ,s}  	
& = &
\frac{-\pi Z}{\sqrt{2^3|E(u_r) |r_r^3}} \left\{\frac{3x_r}{r_r}  P_{-\frac{3}{2}}^{-1}\left(-\sign{[E(u_r)]}\frac{ x_r}{r_r}\right) \right.\nonumber\\&&\left. - \sign{[E(u_r)]} \;P_{-\frac{1}{2}}^{-1}\left(-\sign{[E(u_r)] }\frac{x_r}{r_r}\right) \right\},
\label{eq:14:corr_peak_preciser_generealized_long1}\\
p_{1 y ,s} 
 	& = & -\frac{Zy_{r}}{\sqrt{y_r^2 + z_r^2}}\frac{3\pi}{\sqrt{2^3|E(u_r)|r_r^3}} \;P_{-\frac{3}{2}}^{-1}\left(\sign{[E(u_r)]}\frac{x_r}{r_r}\right),\label{eq:14:corr_peak_preciser_generealized_y}\\
 	p_{1 z ,s} 
 	& = & -\frac{Zz_{r}}{\sqrt{y_r^2 + z_r^2}}\frac{3\pi}{\sqrt{2^3|E(u_r)|r_r^3}} \;P_{-\frac{3}{2}}^{-1}\left(\sign{[E(u_r)]}\frac{x_r}{r_r}\right),
\label{eq:14:corr_peak_preciser_generealized_trans1}
\end{eqnarray}
where  $r_r=\sqrt{x_r^2+y_r^2+z_r^2}$ and $P_{\nu}^{\mu}(\eta)$ is the Legendre function of the first kind which emerges during the integration as shown in Eq.~(\ref{eq:app:01:10-integral_legendre}).

For illustration, we show the behavior of the Legendre functions in Fig \ref{fig:08:Legendre}. Both functions diverge for $\eta \rightarrow -1$, which is, however, out of the region of our interest since the condition $x_r \lesssim \sqrt{y_r^2 + z_r^2}$ is fulfilled at the recollision, and leads to the restriction on the argument of the Legendre function $\left|\sign{[E(u_r)]}\frac{x_r}{r_r}\right| \lesssim \frac{1}{\sqrt{2}}$. The other possible case of $x_r \gg \sqrt{y_r^2 + z_r^2}$ along with $E(u_r)x_r > 0$ does not belong to the soft recollision case. It corresponds to the hard recollision case with a large R-CMT which is beyond the present treatment. \\

In the case of common SR $x_r \rightarrow 0$, we have an approximate formula for R-CMT:
\begin{eqnarray}
p_{1 x,s}(u_r) 	& \approx & 
  Z\frac{\sign{[E(u_r)]}}{3\sqrt{\left|E(u_r)\right|}}\frac{2^{3/2} {\cal P}_1}{r_r^{3/2}}.
\label{eq:13:corr_peak_simplest_generealized_long}\\
p_{1 y,s}(u_r) 		& \approx &  
-Z\frac{2^{3/2} {\cal P}_2y_r}{\sqrt{\left|E(u_r)\right|} r_r^{5/2}}, 
\label{eq:13:corr_peak_simplest_generealized_trans_y}\\
p_{1 z,s}(u_r) 		& \approx &  
-Z\frac{2^{3/2} {\cal P}_2z_r}{\sqrt{\left|E(u_r)\right|} r_r^{5/2}}, 
\label{eq:13:corr_peak_simplest_generealized_trans_z}
\end{eqnarray}
where ${\cal P}_1=3\pi P_{-1/2}^{-1}(0)/8\approx 1.27$, and ${\cal P}_2=3\pi P_{-3/2}^{-1}(0)/8\approx 0.927$ [see Eqs.~(\ref{eq:app:P_three_halfs})-(\ref{eq:app:P_one_halfs})].

\subsection{Fast recollisions}\label{sec:FR}

In the case of  generalized FR we assume $x_r=0$ and relax the condition on acceleration to  $E(u_{r})\approx 0$. In estimating R-CMT, the electron trajectory near the recollision point in the leading order can be then approximated with
\begin{eqnarray}
x_0(u) &\approx& \frac{p_{x r}}{\omega}\sigma,\,\nonumber\\
y_0(u) &\approx&  y_r + \frac{p_{y r}}{\omega}\sigma  \label{eq:FR_leading_order_x_z},\,\\
z_0(u) &\approx&  z_r + \frac{p_{z r}}{\omega}\sigma\nonumber
\end{eqnarray}
which gives  for R-CMT

\begin{widetext}
\begin{eqnarray}
p_{1x, f} &\approx & Z\left[
\frac{
p_{x r}\left[\left(y_r p_{y r} + z_r p_{z r}\right)\sigma + \omega r_r\right]
}{
\left[p_{x r}^2 r_r^2 +\left(y_r p_{z r} - z_r p_{y r}\right)^2 \right]\sqrt{p_{x r}^2\sigma^2 + \left(p_{y r}\sigma+ \omega y_r\right)^2+ \left(p_{z r}\sigma+ \omega z_r\right)^2}}
\right]_{\sigma_1}^{\sigma_2},
\label{eq:10:px_corr_plateau_preciser_TP_asy_x_non-dipole}\\
p_{1y, f} & \approx & Z\left[
\frac{
-y_r p_{x r}^2 \sigma - \left(y_r p_{z r} - z_r p_{y r}\right) \left(\omega z_{r\phantom{y}\!\!\!} + p_{z r} \sigma\right)
}{\left[p_{x r}^2 r_r^2 +\left(y_r p_{z r} - z_r p_{y r}\right)^2 \right]\sqrt{p_{x r}^2\sigma^2 + \left(p_{y r}\sigma+ \omega y_r\right)^2+ \left(p_{z r}\sigma+ \omega z_r\right)^2}}
\right]_{\sigma_1}^{\sigma_2},
\label{eq:10:pz_corr_plateau_preciser_TP_asy_y_non-dipole}\\ 
p_{1 z, f} & \approx &	Z\left[
\frac{
-z_r p_{x r}^2 \sigma - \left(y_r p_{z r} - z_r p_{y r}\right) \left(\omega y_{r \phantom{y}\!\!\!} + p_{y r} \sigma\right)
}{\left[p_{x r}^2 r_r^2 +\left(y_r p_{z r} - z_r p_{y r}\right)^2 \right]\sqrt{p_{x r}^2\sigma^2 + \left(p_{y r}\sigma+ \omega y_r\right)^2+ \left(p_{z r}\sigma+ \omega z_r\right)^2}}
\right]_{\sigma_1}^{\sigma_2}.
\label{eq:10:pz_corr_plateau_preciser_TP_asy_z_non-dipole}
\end{eqnarray}
\end{widetext}
While in the dipole limit relation $y_r p_{z r} = z_r p_{y r}$ holds due to symmetry, in the non-dipole case  $y_r p_{z r} - z_r p_{y r} \approx  y_r [p_{zd}(u_r,u_i) - \frac{1}{u_r - u_i}\int_{u_i}^{u_r}p_{zd}(u,u_i) du]\sim y_r \,c\xi^2$. The terms proportional to $(y_r p_{z r} - z_r p_{y r} )^2$ in the denominators in Eqs.~(\ref{eq:10:px_corr_plateau_preciser_TP_asy_x_non-dipole})-(\ref{eq:10:pz_corr_plateau_preciser_TP_asy_z_non-dipole}) are $\xi^2$ times smaller with respect to the leading term [in estimation we use $p_{xr}\sim c\xi$] 
and can be therefore neglected with respect to the expansion parameter $\epsilon$ of Eq.~(\ref{d-rho}) which can be rewritten as $\epsilon \sim  (\xi/\gamma)\sqrt{ E_a/E_0}$. Using $p_{xr}\sigma /\omega \sim y_r$, one can show negligible contributions of the magnetic drift terms with respect to the leading one in numerators of Eqs.~(\ref{eq:10:pz_corr_plateau_preciser_TP_asy_y_non-dipole}) and (\ref{eq:10:pz_corr_plateau_preciser_TP_asy_z_non-dipole}) by a factor of  $\xi$.


 Thus, after neglecting the discussed terms, we have
\begin{eqnarray}
p_{1x, f} &\approx &
Z\left[\frac{p_{\perp r}\sigma + \omega }{p_{x r}r_r\sqrt{p_{x r}^2\sigma^2 + \left(p_{y r}\sigma+ \omega y_r\right)^2+ \left(p_{z r}\sigma+ \omega z_r\right)^2}}
\right]_{\sigma_1}^{\sigma_2}, \nonumber\\ 
\label{eq:11:px_corr_plateau_preciser_TP_asy_x}\\
p_{1y, f} & \approx &	
Z\left[\frac{- \sigma y_r}{r_r^2\sqrt{p_{x r}^2\sigma^2 + \left(p_{y r}\sigma+ \omega y_r\right)^2+ \left(p_{z r}\sigma+ \omega z_r\right)^2} }\right]_{\sigma_1}^{\sigma_2},
\label{eq:11:pz_corr_plateau_preciser_TP_asy_y}\\ 
p_{1 z, f} & \approx &	
Z\left[\frac{- \sigma z_r}{r_r^2\sqrt{p_{x r}^2\sigma^2 + \left(p_{y r}\sigma+ \omega y_r\right)^2+ \left(p_{z r}\sigma+ \omega z_r\right)^2}}\right]_{\sigma_1}^{\sigma_2},
\label{eq:11:pz_corr_plateau_preciser_TP_asy_z}
\end{eqnarray}
with $\sigma_1$ and $\sigma_2$ being the lower and upper limits of integration, respectively and $p_{\perp r} = \sqrt{p_{y r}^2+p_{z r}^2}$. 

In the rescattering picture, the limits $\sigma_1$ and $\sigma_2$ can be set to $\pm\infty$,  yielding for FR
\begin{eqnarray}
p_{1x,f} 	& \approx &  \frac{2 Z p_{\perp r}}{r_r p_{x r} p_r },
 \label{eq:12:corr_plateau_simplest_generealized_long1}\\
p_{1y,f} 		& \approx &  -\frac{2 Z y_r}{r_r p_r },
 \label{eq:12:corr_plateau_simplest_generealized_trans1_y}\\
 p_{1z,f} 		& \approx &  -\frac{2 Z z_r}{r_r p_r},
 \label{eq:12:corr_plateau_simplest_generealized_trans1_z}
\end{eqnarray}
where $p_{r}=\sqrt{p_{x r}^2+p_{yr}^2+p_{zr}^2}$.

For high-order FR, the recollision picture may break down, which means that the Coulomb momentum transfer (although being rather small) is not decreasing sharply when the electron leaves the recollision point. In this case the Eqs. (\ref{eq:12:corr_plateau_simplest_generealized_long1})-(\ref{eq:12:corr_plateau_simplest_generealized_trans1_z}) provide not a good approximation. Our analysis  shows that better approximation is achieved with $\sigma_1 = - \text{Mod}(u_r,\pi)$ and $\sigma_2 = \pi - \text{Mod}(u_r,\pi)$, which corresponds to the integration between the surrounding turning points of the trajectory.  For FR (i.e., with vanishing laser field  $E(u_r) \approx 0$) and beyond the recollision picture, we can set $\sigma_1 = -\sigma_2 = -\pi/2$  in Eqs. (\ref{eq:11:px_corr_plateau_preciser_TP_asy_x})-
(\ref{eq:11:pz_corr_plateau_preciser_TP_asy_z}) yielding for FR in the leading term
\begin{eqnarray}
p_{1 x, fb} &\approx &
\frac{2\pi Z p_{\perp r}}{p_{x r}r_r\sqrt{p_r^2\pi^2 + 4\omega^2 r_r^2}},
\label{eq:11:px_corr_plateau_preciser_turnpoins}\\
p_{1 y, fb} & \approx &	
-\frac{2\pi Z _r}{r_r^2\sqrt{p_r^2\pi^2 + 4\omega^2 r_r^2}}, 
\label{eq:11:py_corr_plateau_preciser_turnpoins}\\ 
p_{1 z, fb} & \approx &	
-\frac{2\pi Z z_r}{r_r^2\sqrt{p_r^2\pi^2 + 4\omega^2 r_r^2}}. 
\label{eq:11:pz_corr_plateau_preciser_turnpoins} 
\end{eqnarray}
Let us estimate conditions when the recollision picture is violated. 
This is the case once the transversal distance at FR is comparable or greater than the amplitude of the quiver motion. There is also a restriction on the longitudinal momentum $p_{x r}$, which can be derived assuming Eqs. (\ref{eq:13:corr_peak_simplest_generealized_trans_y})-(\ref{eq:13:corr_peak_simplest_generealized_trans_z}) and (\ref{eq:12:corr_plateau_simplest_generealized_trans1_y})-(\ref{eq:12:corr_plateau_simplest_generealized_trans1_z}) to yield comparable results R-CMT. These conditions read
\begin{eqnarray}
r_r	&  \gtrsim & \frac{|E(u_r)|}{\omega^2}, 
\label{plateauConditionForSimplest_zr}\\
p_{x r} 	&  \lesssim   & \sqrt{\frac{|E(u_r)|r_r}{2}}.
\label{plateauConditionForSimplest_pxr}
\end{eqnarray}

The R-CMT formulas for SR Eqs.~(\ref{eq:14:corr_peak_preciser_generealized_long1})-(\ref{eq:14:corr_peak_preciser_generealized_trans1})[or simplified Eqs.~(\ref{eq:13:corr_peak_simplest_generealized_long})-(\ref{eq:13:corr_peak_simplest_generealized_trans_z})], and for FR Eqs.~(\ref{eq:12:corr_plateau_simplest_generealized_long1})-(\ref{eq:12:corr_plateau_simplest_generealized_trans1_z}) [or Eqs.~(\ref{eq:11:px_corr_plateau_preciser_turnpoins})-(\ref{eq:11:pz_corr_plateau_preciser_turnpoins}) when the recollision picture fails]
are expressed via parameters (coordinate and momentum) of the recollision  and are valid even in the case when the global electron trajectory is significantly disturbed by the Coulomb field  with respect to the laser driven one. 
The non-dipole effects in these formulas are accounted for in the parameters $z_r$ and $p_{z r}$.

\subsection{Simpleman estimations}
\label{subsec:simplemans}

The leading scaling of R-CMT  in Eqs.~(\ref{eq:13:corr_peak_simplest_generealized_long})-(\ref{eq:13:corr_peak_simplest_generealized_trans_z}),  (\ref{eq:12:corr_plateau_simplest_generealized_long1})-(\ref{eq:12:corr_plateau_simplest_generealized_trans1_z}) 
can be explained from the following intuitive consideration. The transverse R-CMT can be estimated as the transversal force $F_{\perp r}\sim 1/r_r^2$ acting during the recollision as
\begin{eqnarray}
p_{1\perp} \sim F_{\perp r}\tau_r,
\label{p1perp}
\end{eqnarray}
where $\tau_r$ is the 
duration of the recollision. We define the half of the recollision duration as a time when the electron longitudinal distance from the core reaches the value of the recollision distance, i.e., $x(\tau_r/2)=r_r$. In the case of SR $x(t)\approx -E(u_r)t^2/2$, and 
\begin{eqnarray}
\tau_{r,s}\sim 2\sqrt{2r_r/|E(u_r)|},
\label{trpeak}
\end{eqnarray}
while for FR $x_{F}(t)\approx p_{x r}t$, and 
\begin{eqnarray}
\tau_{r,f}\sim  2z_r/p_{x r}.
\label{trplateau}
\end{eqnarray}
Thus, from Eqs.~(\ref{p1perp})-(\ref{trplateau}), we find estimations 
for the transverse R-CMT,
\begin{eqnarray}
p_{1\perp ,s}  &\sim & -\frac{2^{3/2}Z}{r_r^{3/2}\sqrt{|E(u_r)|}},\label{p1perppeak}\\
p_{1\perp ,f}  &\sim & -\frac{2Z}{r_r p_{x r} }.
\label{p1perpplateau}
\end{eqnarray}
The longitudinal R-CMT at SR is easily estimated from the longitudinal force $F_{\parallel\, r} \sim -x_r(t)/z_r^3$ via
\begin{eqnarray}
p_{1 \parallel, s}  \sim \int^{\tau_{r,s}/2}_{-\tau_{r,s}/2} F_{\parallel\, r}dt \sim -\frac{2 Z}{r_r^3}\int_0^{\tau_{r,s}/2}x_s(t)dt=\frac{Z E(u_r)\tau_{r,s}^3}{2^33r_r^3},\nonumber
\end{eqnarray}
which yields
\begin{eqnarray}
p_{1\parallel, s}  \sim  \frac{2^{3/2} Z}{3\sqrt{|E(u_r)|}r_r^{3/2}}.
\label{p1longpeak}
\end{eqnarray}
For estimation of the longitudinal R-CMT at the FR one has to take into account that there is a compensation of R-CMT stemming from trajectories before and after the recollision which can be incorporated by a re-establishment of the time-dependence of $z_r(\tau) = z_r + p_{\perp r}\tau$ in $F_{\parallel\, r}$ and in the limit of $p_{\parallel r} \gg p_{\perp r}$:
 \begin{eqnarray}
p_{1\parallel,f}  \sim -Z\int_0^ {\tau_{r,f}/2 } \left[\frac{p_{x r}\tau}{(z_r+p_{\perp r}\tau)^3}- \frac{p_{x r}\tau}{(z_r-p_{\perp r}\tau)^3}\right]d\tau \approx   \frac{Z p_{x r}p_{\perp r} \tau_{r,f}^3}{2^2z_r^4}.\nonumber
\label{p1long}
\end{eqnarray}
Once substituted from Eq.~(\ref{trplateau}), we obtain the final formula
\begin{eqnarray}
p_{1\parallel,f}  \sim  \frac{2 Z p_{\perp r} }{z_rp_{x r}^2 }.
\label{p1longplateau}
\end{eqnarray}
Thus, by the applied simple estimations the leading scaling of R-CMT from Eqs.~(\ref{eq:12:corr_plateau_simplest_generealized_long1})-(\ref{eq:12:corr_plateau_simplest_generealized_trans1_z}) are reproduced  in the limit of $p_{\perp r} \ll p_{x r}$.  


\subsection{Global perturbation approach for the Coulomb field}
\label{subsec:dipole:characteristic_trajectries}

Formulas for R-CMT in Subsecs.~\ref{sec:SR} and \ref{sec:FR} depend on the recollision parameters, momentum and coordinate. These parameters can be derived explicitly when one adopts global perturbation approach for the Coulomb field of the atomic core. In this approach the analytical description of the recolliding trajectories are provided in Sec.~\ref{sec:dipole:03:distinctive_traj}.  
We have defined two types of characteristic recolliding trajectories along with analytical description of the recolliding and ionization phase. 
Once inserted into the general formulas, we can investigate the direct scaling of R-CMT with respect to the laser parameters for those trajectories. For simplicity we will assume in this subsection dipole approximation and will set $p_{y i} = 0$ and $p_{z i} = p_{\bot i}$.

By employing $ y_r = 0$ and $z_r=p_{\perp i}(u_r-u_i)/\omega$, the relation between the ionization and recollision phases $u_r$  given by Eq.~(\ref{eq:04:00:peak_ur}), and $u_i$ by Eq.~(\ref{eq:04:peak_ui}) into the Eqs.~(\ref{eq:13:corr_peak_simplest_generealized_long})-(\ref{eq:13:corr_peak_simplest_generealized_trans_z}), we obtain for SR:
\begin{eqnarray}
p_{1 \parallel, s}^{(k)} &=&  Z\frac{(-1)^{k+1}{\cal P}_1}{3\sqrt{E_0}}
  \left[\frac{2\omega}{p_{\perp i}(k + 1)\pi}\right]^{3/2},
\label{eq:12_px_corr_peak_preciser_alpha_infinity}\\
p_{1\bot, s}^{(k)} &=&	 -  Z \frac{{\cal P}_2}{\sqrt{E_0}}\left[\frac{2\omega}{p_{\perp i} (k + 1)\pi}\right]^{3/2}.\label{eq:12_pz_corr_peak_preciser_alpha_infinity}
\end{eqnarray}

At usual FR, one has $p_{x r}\gg p_{\perp r}$. The zero-order  laser driven trajectory is $p_{xr}=A_x(u_r)-A_x(u_i)$. For  FR using $u_r$  from Eq.~(\ref{eq:04:00:plateau_ur}), and $u_i$ by Eq.~(\ref{eq:04:plateau_ui}), in the formulas of Eqs.~(\ref{eq:12:corr_plateau_simplest_generealized_long1})-(\ref{eq:12:corr_plateau_simplest_generealized_trans1_z}), we have 
\begin{eqnarray}
p_{1 \parallel, f}^{(l)} &\approx&	 Z\frac{4(-1)^{l+1}\omega^3}{E_0^2\left[{\left(\frac{2}{\pi(2l+1)}\right)-(-1)^{l}}\right]^2\pi\left(2l+1\right)},
\label{eq:11:px_corr_plateau_preciser}\\
p_{1\bot, f}^{(l)} &\approx&	-Z \frac{4\omega^2}{E_0 p_{\perp i}\left|{\left(\frac{2}{\pi(2l+1)}\right)-(-1)^l}\right|\pi\left(2l+1\right)}.
\label{eq:11:pz_corr_plateau_preciser}
\end{eqnarray}

For the FR, when  the recollision picture begins to break down, we derive from Eqs.~(\ref{eq:11:px_corr_plateau_preciser_turnpoins})-(\ref{eq:11:pz_corr_plateau_preciser_turnpoins}):
\begin{eqnarray}
p_{1{ \parallel}, fb}^{(l)} &=&  \frac{4 Z \omega^3 \left(\frac{2}{\pi(2l+1)} - (-1)^l\right) ^{-1}}{E_0^2\pi(2l+1)\sqrt{\left(\frac{2}{\pi(2l+1)} - (-1)^l\right)^2 + p_{\perp i}^2\frac{\omega^2}{E_0^2}(2l+1)^2}},\nonumber\\
\label{eq:12_px_corr_peak_preciser_turnpoints}\\
 p_{1\bot, fb}^{(l)} &=&	 -\frac{4 Z \omega^2}{E_0 p_{\perp i}\pi(2l+1)\sqrt{\left(\frac{2}{\pi(2l+1)} - (-1)^l\right)^2 + p_{\perp i}^2\frac{\omega^2}{E_0^2}(2l+1)^2}}.\nonumber\\
\label{eq:12_pz_corr_peak_preciser_turnpoints}
\end{eqnarray}

The R-CMT estimations of this subsection are applicable in the case when the Coulomb field treated as a global perturbation. They provide us with the scaling of the R-CMT with respect to the laser parameters,  the electron initial transverse momentum, and the order of the rescattering $k$ or $l$.

We analyse the accuracy of the global approach in Appendix~\ref{sec:app:accuracy-R}.
The simple estimates of Eqs.~(\ref{eq:11:px_corr_plateau_preciser})-(\ref{eq:11:pz_corr_plateau_preciser}) are more accurate
for several first rescatterings (up to the 5th for transversal and
3th for the longitudinal) where the recollision picture holds.
The formulas (\ref{eq:12_px_corr_peak_preciser_turnpoints})-(\ref{eq:12_pz_corr_peak_preciser_turnpoints}) deliver better estimations for larger $k$ and $l$ which is connected to the break-down of the recollision picture and integration within the interval given by the closes turning points is more suitable. The estimations demonstrate nontrivial suppression of the R-CMT with increasing order of rescattering $k$ or $l$.

In this section we have derived formulas for estimation of R-CMT for specific recollision events, SR and FR. The transverse R-CMT for both of the recollisions can be represented in a unified form
\begin{eqnarray}
\textbf{p}_{1\bot}\approx -2 Z \tau_r\frac{ \textbf{r}_r}{r_r^3},\label{p-bot}
\end{eqnarray}
with an appropriate time of recollision $\tau_r$, different for SR and FR.
Using these formulas,  further we will  estimate the total R-CMT for an arbitrary trajectory. However, before the implementation of this task, we need  an estimate for I-CMT as accurate as that of R-CMT, which is carried out in the next section. The accuracy of these expressions in the dipole approximation case are analysed in Appendix \ref{sec:app:accuracy-R}, where a comparison with the exact numerical results is given.

\section{Initial Coulomb momentum transfer}\label{sec:dipole:05:intial_mom_transfer}

For the analytical estimation of I-CMT, we have to calculate the Coulomb  momentum transfer to the electron which takes place immediately after the leaving the tunnel exit by using Eq.~(\ref{eq:01:03:first_order:momentum_trans}). The electron is at the tunnel exit $x_i$ at ionization phase $u_i$ with a transversal momentum $p_{\bot i}$ and is furher accelerated by the laser field $E(u_i)$ in longitudinal direction. We assume that the transversal motion is much smaller than the longitudinal one and expand the denominator of Eq. (\ref{eq:01:03:first_order:momentum_trans}):
 \begin{eqnarray}
\frac{1}{\left[x^2(u) + y^2(u)+z^2(u)\right]^{3/2}} &\approx& \frac{1}{|x(u)|^{3}}\left[
1 - \frac{3}{2}\frac{y^2(u)+z^2(u)}{x^2(u)} \right].\nonumber\\
\label{eq:10:init_dist_expansion}
\end{eqnarray}
Taking into account that  $y=p_{yi}\sigma /\omega$ and $z=(p_{zi}+\overline{p}_{zd})\sigma /\omega$, the second term in the bracket can be estimated as
\begin{eqnarray}
\frac{y^2(u)+z^2(u)}{x^2(u)}\sim \frac{p_{\bot i}^2\sigma^2}{\left(x_i-\frac{E_0}{2\omega^2}\sigma^2\right)^2\omega^2}+\frac{2p_{\bot i}\overline{p}_{zd}\sigma^2}{\left(x_i-\frac{E_0}{2\omega^2}\sigma^2\right)^2\omega^2}.\label{transverse_expansion}
\end{eqnarray}
The first term in Eq.~(\ref{transverse_expansion}) is dominant over the second one by a factor of $\epsilon$. The order of magnitude of the first term is $\sim E_0/E_a\ll 1$, which justifies the expansion above. We estimated
 the effective region of $\sigma\equiv u-u_i$  from the relation $E_0\sigma^2/\omega^2\sim x_i$. 


\subsection{The first-order approximation}

The first order approximation for I-CMT uses the unperturbed trajectory 
\begin{eqnarray}
x_{0}(\sigma) & \approx & x_i - \frac{E(u_i)}{2\omega^2}\sigma^2,
\label{eq:10:init_corr_coordinates_long_SS} \\
z_{0}(\sigma) & \approx & \frac{p_{\perp i}}{\omega}\sigma,
\label{eq:10:init_corr_coordinates_trans_SS}
\end{eqnarray}
with the coordinate $z_0(\sigma)$ along the initial transverse momentum, and the momentum corrections in this order are  
\begin{eqnarray}
p_{1 \parallel} (u) & = & -\frac{Z}{\omega}\int\limits_{u_i}^{u}
\frac{\sign{(x_{0}(u))}}{x_{0}^{2}(u)}
\dd u,
\label{eq:12:init_p1x_def}\\
p_{1 \perp} (u) & = & -\frac{Z}{\omega}\int\limits_{u_i}^{u}
\frac{z_{0}(u)}{|x_{0}^{3}(u)|}
\dd u ,
\label{eq:12:init_p1z_def}
\end{eqnarray}
The integrals can be easily evaluated for $x_{i} < 0$ as
\begin{eqnarray}
p_{1\parallel}(u) & = &
\frac{Z \sigma}{2x_i^2 \omega \left[1 + \sigma^2/\gamma^2(u_i)\right]}+
\frac{Z \gamma (u_i)\arctan{\left[\sigma/\gamma (u_i)\right]}}{2x_i^2\omega}
,\nonumber\\
p_{1\perp}(u) & = &\frac{Z p_{\perp i}}{2E(u_i)x_i^2}\left\{\frac{1}{\left[1 + \sigma^2/\gamma^2(u_i)\right]^2 } -1\right\},
\label{eq:12:init_p1z_solved}
\end{eqnarray}
with $\gamma (u_i)=\sqrt{2I_p}\omega/|E(u_i)|$, which at $\sigma \rightarrow \infty$ yields I-CMT formulas as derived in \cite{Shvetsov-Shilovski_2009} reading
\begin{eqnarray}
p_{1\parallel, \text{in}} & = &\frac{Z \pi }{\sqrt{2^3|x_i|^3|E(u_i)|}}=\frac{Z \pi E(u_i)}{(2I_p)^{3/2}},\label{ICMTt}\\
p_{1\perp,\text{in}} & = &-\frac{Z p_{\perp}}{2x_i^2|E(u_i)|}=-\frac{2Z p_\perp |E(u_i)|}{(2I_p)^{2}},\label{ICMTl}
\end{eqnarray}

\subsection{Second-order corrections}

For calculation of the second-order I-CMT we need the first-order correction to the trajectory, which is found integrating Eqs.~(\ref{eq:12:init_p1z_solved}):
\begin{eqnarray}
x_{1}(\sigma) & = & \frac{Z\sigma\gamma (u_i)}{2x_i^2   \omega^2 }
\arctan{\left(\frac{\sigma}{\gamma (u_i)}\right)}  \approx \frac{Z\sigma^2}{2\omega^2 x_{i}^{2}} ,
\label{eq:13:init_x1_solved}\\
z_{1} (\sigma) 
		 & \approx & -\frac{Zp_{\perp i}\sigma^3}{6\omega^3 |x_i^3|},
\label{eq:13:init_z1_solved}
\end{eqnarray}
where we keep only the leading terms in the expansion over the small  $\sigma$. 
The correction to I-CMT is calculated using the first-order correction to the $x$-coordinate of the trajectory, but neglecting the correction to the $z$-coordinate, as it is small, determined by the small initial transverse momentum.

Combining the first- and the second-order momentum corrections, and expanding over the small parameter $1/(|E(u_i)| x_i^2 )\sim E_0/E_a$,  we arrive at the following expressions for the corrected I-CMT:
\begin{eqnarray}
p_{2\parallel,\text{in}} & = & \frac{\pi Z \sign{(E(u_i))}}{\sqrt{2^3 |E(u_i) x_i^3|}}\left[1 + \frac{4Z - 3p_{\perp i}^2|x_i|}{8|E(u_i)| x_i^2}  + \mathcal{O}\left(\frac{1}{x_i^{4} E_0^{2}}\right)\right],\nonumber\\
\label{correctedICMTxit}\\
p_{2\perp,\text{in}} & = & 
-\frac{Z p_{\perp i}}{2 |E(u_i)| x_i^2} \left[1  + \frac{4Z-3p_{\perp i}^2|x_i|}{6|E(u_i)| x_i^2}  + \mathcal{O}\left(\frac{1}{x_i^{4}E_0^{2}}\right)\right],
\label{correctedICMTxil}
\end{eqnarray}
which read in the quasistatic regime with $x_i = -I_p/E(u_i)$:
\begin{eqnarray}
p_{2\parallel,\text{in}} & = & \frac{Z \pi E(u_i)}{(2I_p)^{3/2}}\left[1+\frac{2Z|E(u_i)|}{E_a\sqrt{2I_p}}-\frac{3p_{\perp i}^2}{8 I_p}  + \mathcal{O}\left(\frac{E_0^2}{E_a^2}\right)\right], \label{correctedICMTt}\\
p_{2\perp,\text{in}} & = & -\frac{2 Z p_{\perp i} |E(u_i)|}{(2I_p)^{2}}\left[1+\frac{8Z|E(u_i)|}{3E_a\sqrt{2I_p}}-\frac{p_{\perp i}^2}{2 I_p}  + \mathcal{O}\left(\frac{E_0^2}{E_a^2}\right)\right].\nonumber\\
\label{correctedICMTl}
\end{eqnarray}

Let us note that it is easy to identify the term $\sim Z$ in the expansions in Eqs.~(\ref{correctedICMTxit})-(\ref{correctedICMTl}) as the second-order momentum correction whereas the term $\sim {p_{\perp i}^2}$ as the correction in the first-order due to the transversal motion.

We compare our results for I-CMT with numerical calculation results in Appendix \ref{App:accuracy-I-CMT}. As we can see the next-to-leading order corrections to I-CMT significantly decrease the error of the estimation.

\section{Total momentum transfer}\label{sec:dipole:07:total}

In this section we illustrate the capability of our analytical approach, providing an estimation of the final momenta for any arbitrary ionization phase $u_i$, and initial transversal momentum $p_{\perp i}$ and, accordingly, the estimation for the asymptotic PMD. This is a good check whether the derived analytical formulas for R-CMT and I-CMT can provide physically relevant results and can help to gain greater insight into the evolution of the tunneled wave packet in the continuum.
Since the evolution of the electrons is nontrivial in the combined laser and Coulomb field, we consider two different approaches: the straightforward zero-order trajectory approach, and the more elaborate step-by-step approach.
Let us note at this point, that for simplicity the dipole approximation is applied and $Z =1$ throughout this section.

\subsection{Zero-order trajectory approach}

\begin{figure}[b]
	\centering
  		  		\includegraphics[width=1\linewidth]{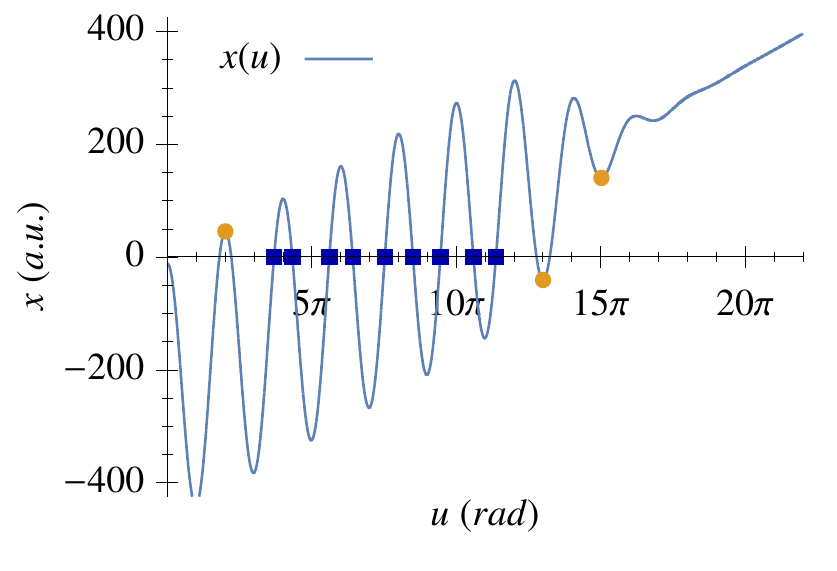}
	    \caption{An illustrative trajectory demonstrating two different types of recollisions: SR (by orange dots) and FR (by blue squares). We keep the SR points if $|x_r| < E_0/(5\omega^2)$ for $x_r\ddot{x}_r < 0$, otherwise we find the closest FR points. We also keep those SR with $x_r\ddot{x}_r > 0$ but neglect any SR points with $\ddot{x}_r < E_0/10$.}
	        \label{fig:15:rescattiring_points}
\end{figure}

In this approach, the Coulomb field of the atomic core is treated as a global perturbation. Then, the zero-order trajectory describes the trajectory of the electron solely in the laser field. The total Coulomb effect mostly amounts to I-CMT and R-CMT (there is also a small asymptotic contribution after the laser pulse is switched off, which will be discussed below). For estimation of R-CMT  we use the zero-order trajectory, find the rescattering points and for each rescattering event apply our R-CMT formulas derived in the orevious sections.  The  I-CMT  distorts the zero-order trajectory significantly which can have an essential impact on the rescattering points. Therefore, we include the I-CMT in the zero-order trajectory via modification of the initial momentum:
\begin{eqnarray}
\tilde{x}_0(u) 	& = &  \frac{E_0}{\omega^2}\left[
\cos{u} - \cos{u_i} + (u - u_i) \left(\sin{u_i}\right)\right] + (u - u_i) p_{2\parallel,\text{in}} +x_i,\nonumber\\ 
\label{eq:13:coordinate_init_distorted_x}\\
\tilde{y}_0(u) 	& = & \frac{p_{yi} + p_{2y, \text{in}}  }{\omega}(u-u_i),\label{eq:13:coordinate_init_distorted_y}\\
\tilde{z}_0(u) 	& = & \frac{p_{zi} + p_{2z, \text{in}}}{\omega}(u-u_i),  
\label{eq:13:coordinate_init_distorted_z}
\end{eqnarray}
with $p_{2y, \text{in}}$ and $p_{2z, \text{in}}$ being projections of the transversal I-CMT $p_{2\perp,\text{in}}\left(u_i, \sqrt{p_{yi}^2+p_{zi}^2}\right)$ on $y-$ and $z-$direction, respectively. 

The final momentum is obtained by including the contribution of all R-CMT into the momentum transfer, yielding 
\begin{eqnarray}
\textbf{p}(u_i,\textbf{p}_{\bot i}) = \textbf{p}_0(u_i,\textbf{p}_{\bot i})+\textbf{p}_{2,\text{in}}(u_i,\textbf{p}_{\bot i})+\sum\limits_{j=1}^{N} \textbf{p}_{1 }^{(j)}(u_i,\widetilde{\textbf{p}}_{\bot i}),\nonumber\\
\label{13:Zero:final_mom_px}
\end{eqnarray}
where $N$ is the total number of effective rescatterings, $\textbf{p}_{0}(u_i)=\left( -A(u_i),p_{yi},p_{zi}\right)$ is the zero-order asymptotic momentum, the initial transverse momentum $\textbf{p}_{\bot i}=(0,p_{yi}, p_{zi})$, the initial momentum correction is 
$\textbf{p}_{2,\text{in}}\left(u_i,\textbf{p}_{\bot i}\right) \equiv p_{2\parallel,\text{in}}(u_i, |\textbf{p}_{\bot i}|) \textbf{e} - \left|p_{2\perp,\text{in}}\left(u_i, |\textbf{p}_{\bot i}|\right)\right| \frac{\textbf{p}_{\bot i}}{|\textbf{p}_{\bot i}|}$, the distorted initial transversal momentum  $\widetilde{\textbf{p}}_{\bot i} \equiv \textbf{p}_{\bot i} - \left|p_{2\perp,\text{in}}\left(u_i, |\textbf{p}_{\bot i}|\right)\right|\frac{\textbf{p}_{\bot i}}{|\textbf{p}_{\bot i}|}$, and $\textbf{p}_{1 }^{(j)}(u_i,\widetilde{\textbf{p}}_{\bot i})$ is the R-CMT at $j^{th}$ recollision given by the formulas derived in the previous sections, corresponding to the specific type of this recollision. 
\begin{figure}
	\centering
 \includegraphics[width=1.0\linewidth]{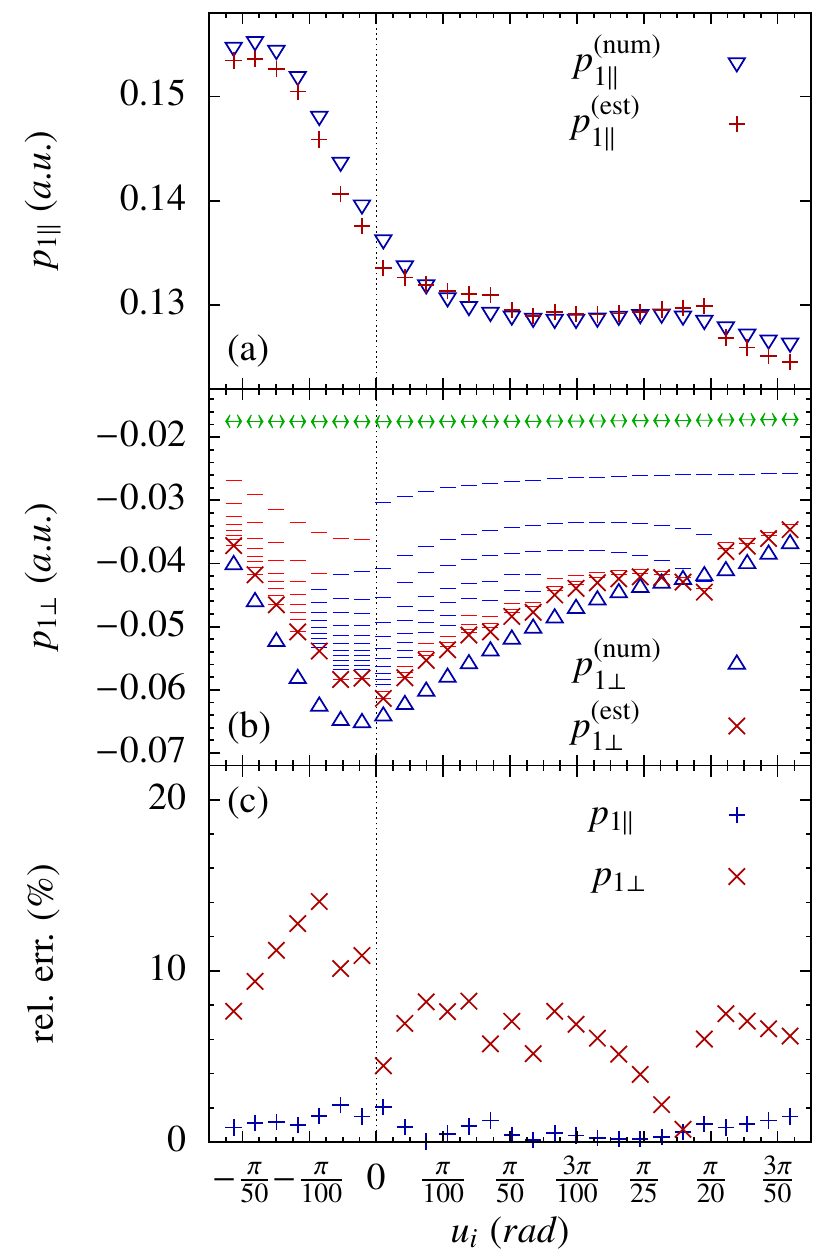}
\caption{Total CMT vs ionization phase $u_i$ for zero-order trajectory: (a) longitudinal momentum transfer; (b) transverse momentum transfer; (c) the corresponding errors. Numerical simulations are shown with triangles, the estimation (see the text) with red crosses,  I-CMT  with green double arrow.  The contribution of each R-CMT is shown by a line segment: SR - red, FR - blue. The contributions are added to  I-CMT and to the previous R-CMTs, as long as it is larger than $5\%$ of total numerical estimate of CMT. This restriction was applied for the sake of graphical simplicity only.}
    \label{fig:13:total_MT_zero_arbitrary_E_0_04}
\end{figure}
\begin{figure}
	\centering
  		  		\includegraphics[width=1.0\linewidth]{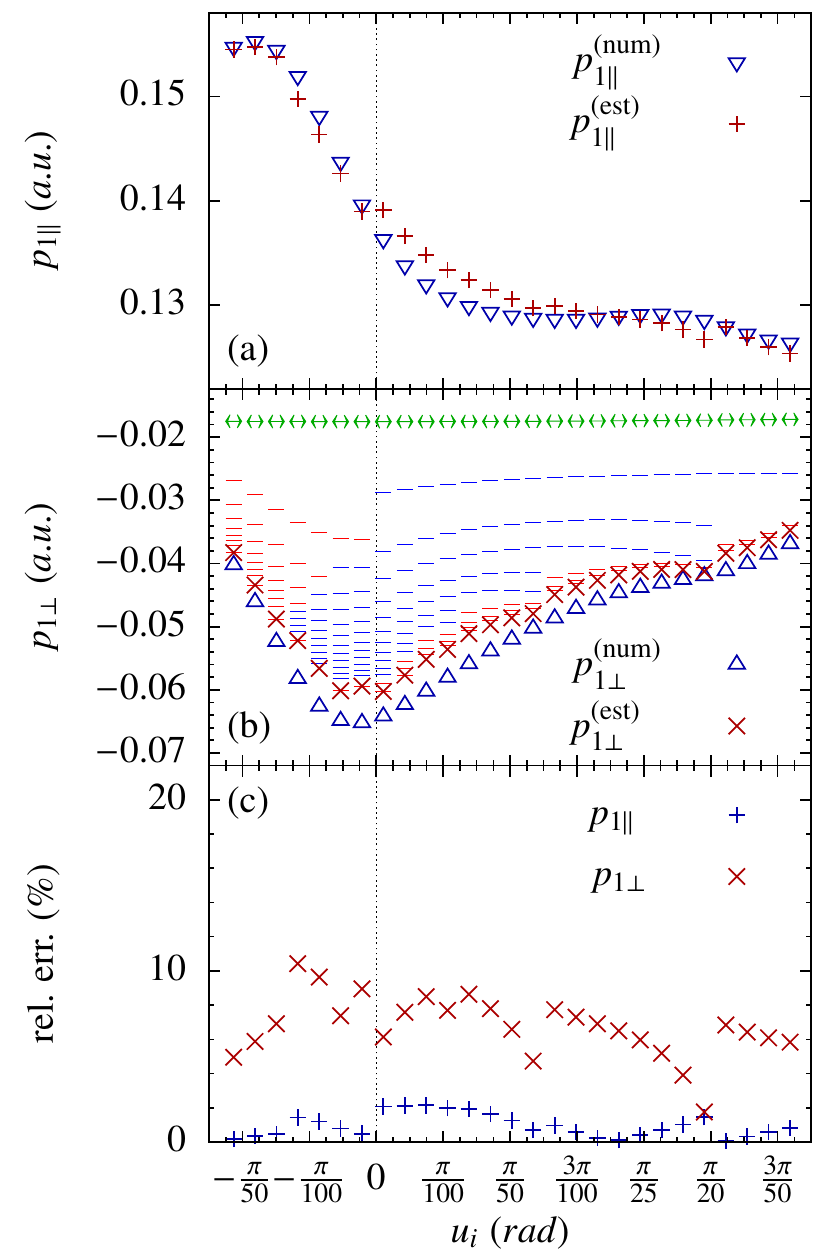}
	    \caption{Total CMT vs ionization phase $u_i$ for the step-by-step evolved  trajectory: (a) longitudinal momentum transfer; (b) transverse momentum transfer; (c) the corresponding errors. Numerical simulations are shown with triangles, the estimation (see the text) with red crosses,  I-CMT  with green double arrow.  The contribution of each R-CMT is shown by a line segment: SR - red, FR - blue. The contributions are added to I-CMT and to the previous R-CMTs, as long as it is larger than $5\%$ of total numerical estimate for CMT. Again, for the sake of clarity only.}
    \label{fig:14:total_mom_corr_Piecewise_arbitrary_E_0_04}
\end{figure}


We treat a recollision as SR, when $\dot{x}_r=0$ and $x_r\ddot{x}_r > 0$, as well as those  when  $\dot{x}_r=0$, $x_r\ddot{x}_r < 0$, but only when $|x_r| < x_{\text{thresh}} = E_0/(5\omega^2)$. For the estimation of R-CMT we use Eqs.~(\ref{eq:14:corr_peak_preciser_generealized_long1})-(\ref{eq:14:corr_peak_preciser_generealized_trans1}) in both cases. All SR with $|\ddot{x}_r|< x_{\text{thresh}} = E_0/(5\omega^2)$ are neglected because such recollisions happen at the end of the laser pulse and have negligible R-CMT, see Fig.~\ref{fig:15:rescattiring_points}. For some electrons this is not true, however most of them are further trapped in the Rydberg states.

Finally, we treat the remaining rescatterings as FR with Eqs.~
(\ref{eq:12:corr_plateau_simplest_generealized_long1})-(\ref{eq:12:corr_plateau_simplest_generealized_trans1_z})
 or Eqs.~(\ref{eq:11:px_corr_plateau_preciser_turnpoins})-(\ref{eq:11:pz_corr_plateau_preciser_turnpoins}), depending on the conditions given by Eqs.~(\ref{plateauConditionForSimplest_zr})-(\ref{plateauConditionForSimplest_pxr}). The only exceptions are the FR closest to any SR with $x_r\ddot{x}_r < 0$ and $|x_r| < x_{\text{thresh}}$, which we neglect since the R-CMT is already taken into account via the SR (see an exemplary SR at $u_r \sim 13\pi$ replacing two closest FR in Fig.~\ref{fig:15:rescattiring_points}).

We compare results of our analytical  estimations with numerical simulations in Fig.~\ref{fig:13:total_MT_zero_arbitrary_E_0_04} (in the latter  a smooth switch-off of the laser pulse is used).
For $u_i > 0$, a very good agreement with the numerical simulations is achieved. The error is well behaved and peaks at the phases where two methods are switching, namely the single SR forks into two FRs. There is an easily understandable discrepancy for $u_i < 0$, since decreasing $u_i$ tilts the electron's quivering trajectory down and hence the first recollision becomes slow which generally yields larger momentum transfer than FR.
For some particularly small and specially negative $u_i$, the momentum transfer is so large that the whole zero-order trajectory is not a valid approximation anymore and our present approach fails. The accuracy issue will be discussed below in Sec.~\ref{sec:PMD}.  

\subsection{Step-by-step}

When the electron is ionized near the peak of the laser field, its drift velocity is small, and rescattering can happen  with a small impact parameter, inducing large distortion of the laser driven trajectory. 
The same can happen when the electron is ionized with a small transverse momentum at other ionization phases.
It is understandable that the zero-order approximation by Eqs. (\ref{eq:13:coordinate_init_distorted_x})-(\ref{eq:13:coordinate_init_distorted_z}) 
fails in this case.  
However, we can improve our estimations by taking into account R-CMT at each recollision during propagation in the laser field. Thus, the electron is propagated by the laser field only step-by-step from the ionization phase $u_i$ till the end of the laser pulse over all the rescattering events, and by correcting the electron momentum by the estimated R-CMT at every single recollision point. This approach is expected to give 
much preciser results, with a wider range of applicability of ionization phases and initial transverse momenta. Moreover, this approach would allow us to incorporate also the, so-called, long trajectories in our model.

Although the laser driven trajectory is disturbed due to R-CMT at the recollisions, the R-CTM itself can be still calculated using perturbation theory because the latter is always applicable at least during the short time of the recollision. This allows us to use the same formulas for R-CMT as in the previous subsection. The only difference is that the zero-order trajectory is replaced by several step-by-step evolved zero-order trajectories.

The drift momentum after the $n^{th}$-recollision is dependent of all R-CMT received at all previous recollisions and can be iteratively defined as 
\begin{eqnarray}
\textbf{p}^{(n)}\left(u_i,\textbf{p}_{\bot i}\right) = \textbf{p}^{(n-1)}\left(u_i,\textbf{p}_{\bot i}\right) + 
\textbf{p}_{1}^{(n)}\left(u_i,\textbf{p}^{(n-1)}\left(u_i,\textbf{p}_{\bot i}\right)\right),\nonumber\\
\label{13:Step:iterative_p}
\end{eqnarray}
where we have for simplicity redefined the I-CMT as zero-order R-CMT: 
$ \textbf{p}_{1}^{(0)}\left(u_i,\textbf{p}_{\bot i}\right) \equiv 
p_{2\parallel,\text{in}}(u_i, |\textbf{p}_{\bot i}|) \textbf{e} - \left|p_{2\perp,\text{in}}\left(u_i, |\textbf{p}_{\bot i}|\right)\right| \frac{\textbf{p}_{\bot i}}{|\textbf{p}_{\bot i}|}$,
 The iteration starts at $n=-1$ with $\textbf{p}^{(-1)}(u_i, p_{\perp i}) := - A_{x}(u_i) \textbf{e} + \textbf{p}_{\perp i}$.
Let us note that $n=0$ corresponds to the momentum after tunneling and before the first rescattering event which happens at $n=1$. The properties of the $n$-th rescattering can be determined from the zero-order trajectory evolved from the $(n-1)^{\text{th}}$ event with the $\mathbf{p}^{(n-1)}(u_i, p_{\perp i})$ momentum yielding
\begin{eqnarray}
	x_{0}^{(n)}(u) &=& \frac{1}{\omega}\int_{u_{in}}^{u} \left(p_{x}^{(n)}(u_i, p_{\perp i}) + A_x(u)\right) du + x_{0}^{(n-1)}(u_{in}),
 \label{13:Step:iterative_x}\\
	y_{0}^{(n)}(u) &=&  p_{yi}^{(n)}(u_i, p_{\perp i})\left(\frac{u - u_{in}}{\omega}\right)  + y_{0}^{(n-1)}(u_{in}),
 \label{13:Step:iterative_y}\\
	z_{0}^{(n)}(u) &=&  p_{zi}^{(n)}(u_i, p_{\perp i})\left(\frac{u - u_{in}}{\omega}\right)  + z_{0}^{(n-1)}(u_{in}),
 \label{13:Step:iterative_z}
\end{eqnarray}
where we set $u_{in} = u_r$ of the $n$-th rescattering. The tunnel exit enters the iteration as $x_{0}^{(-1)} = x_i$, $y_{0}^{(-1)} = z_{0}^{(-1)} = 0$, and for $n=0$ we have $ u_{in} = u_i$.

\begin{figure}
	\centering
 	 		\includegraphics[width=1\linewidth]{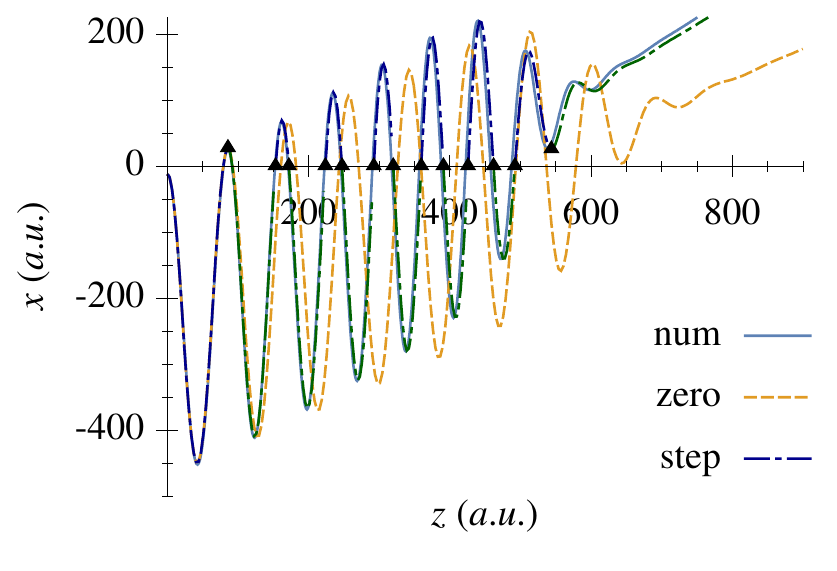}
	    \caption{The electron trajectories in different approaches: (solid, gray) numerical solution, (dashed, orange) zero-order approach, (dot-dashed, multi-color) step-by-step approach.
	    Rescattering points are noted by black triangles. All trajectories match well for the first half-period but start to differ after the first rescaterring event. The used parameters are $E_0 = 0.041$, $\omega = 0.0134$, $u_i = -\pi/188$ and $p_{\perp i} = 0.2$.}
    \label{fig:16:methods_illustrations}
\end{figure}

Trajectories obtained with the step-by-step and the zero-order approaches are compared with the numerical simulation for $u_i = - \pi/100$ in Fig. \ref{fig:16:methods_illustrations}. As we can see, the trajectories do not differ for the first half-period which is achieved by taking the initial momentum correction into account. The difference starts to manifest during the second half-period of the laser field (i.e., after the first rescattering), however, the step-by-step zero-order trajectory approach provides a rather good approximation for the exact numerical trajectory. 

We plotted the resulting CMT for various ionization phases and fixed $p_{\perp i} = 0.2$ a.u. in Fig. \ref{fig:14:total_mom_corr_Piecewise_arbitrary_E_0_04}. The relative error does not change much for the positive phases where only few rescatterings take place. For the negative phases we can actually see an increase in the precision which is a good indication that our step-by-step approach could deliver much better results.

We underline an important message of Figs.~\ref{fig:13:total_MT_zero_arbitrary_E_0_04} and \ref{fig:14:total_mom_corr_Piecewise_arbitrary_E_0_04}, which has been enabled by our analytical approach. A single rescattering is not sufficient to describe the CMT.  The contribution of high-order rescatterings to the total CMT is significant and should not be neglected for a good quantitative  description.

Although the procedure of finding the right rescattering points seems to be straightforward, we need a quite good algorithm selecting them automatically in order to automatize the methods. The algorithms for selection of the proper rescattering point can be found in appendix \ref{App:methods}.

\begin{figure}
	\centering
  		  		\includegraphics[width=1\linewidth]{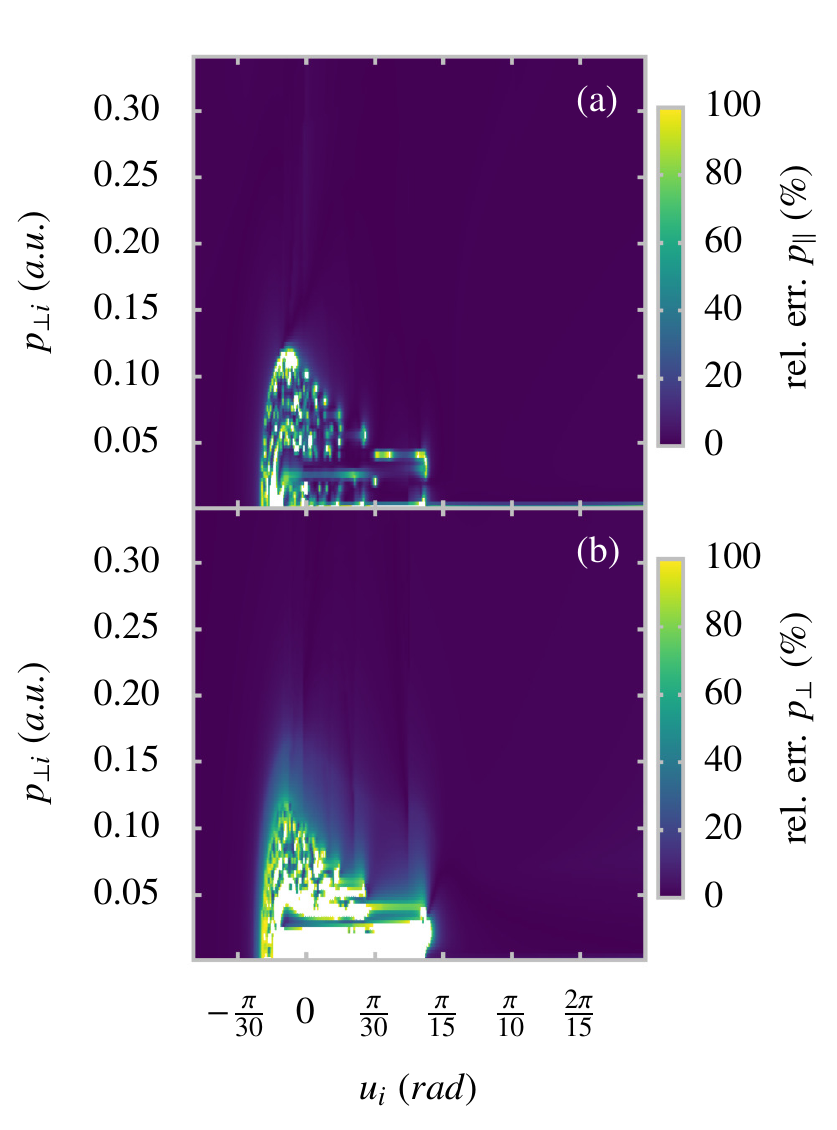}
\caption{Relative error of the photoelectron estimated asymptotic momentum in the  zero-order trajectory approach: (a) longitudinal momentum, (b) transverse momentum, 
(b) longitudinal momentum.}
\label{fig:13:applicability_zero}
\end{figure}
\begin{figure}
	\centering
  		  		\includegraphics[width=1\linewidth]{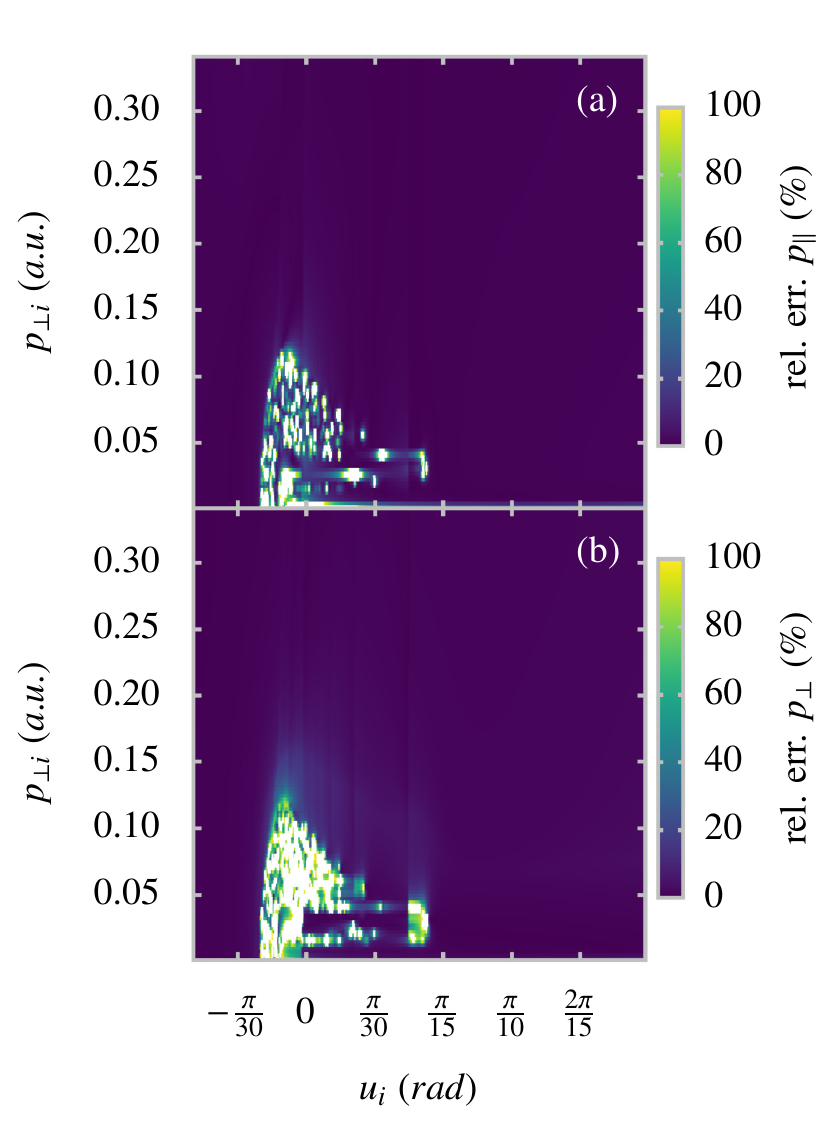}
\caption{Relative error of the photoelectron estimated asymptotic momentum in the  step-by-step approach: (a) longitudinal momentum, (b) transverse momentum.}
\label{fig:18:applicability_step}
\end{figure}

\subsection{Comparing methods}

Let us compare the accuracy of both methods over the whole valid parameter space. 
For a special class of initial conditions, the components of the final momentum can be vanishing, leading to an artificial enhancement of the relative error. Therefore, we re-define the relative error as follows: 
\begin{eqnarray}
\delta p_{\parallel} & = & \left|\frac{p_{\parallel}(u_i, p_{\perp i}) - p_{\parallel}^{(\text{num})}(u_i, p_{\perp i}) }{\max{\left[p_{\parallel}^{}(u_i, p_{\perp i}), p_{\parallel}(u_i, p_{\perp i}) - p_{0\parallel}(u_i)\right]}}\right|100\%, \label{eq:21:errors_px}\\
\delta p_{\perp}		 & = & \left|\frac{p_{\perp}(u_i, p_{\perp i}) - p_{\perp}^{(\text{num})}(u_i, p_{\perp i}) }{\max{\left[p_{\perp}^{}(u_i, p_{\perp i}), p_{\perp}(u_i, p_{\perp i}) - p_{0\perp}(u_i)\right]}}\right|100\%,
\label{eq:21:errors_pz}
\end{eqnarray}
with $p_{\parallel/\perp}\left(u_i, p_{\perp i}\right)$ being the proper component of the electron's final momentum given by Eq.~(\ref{13:Zero:final_mom_px}) or by Eq.~(\ref{13:Step:iterative_p}) at $n = N$ for zero-order or step-by-step method, respectively. The superscript ``$(\text{num})$'' denotes the corresponding value obtained numerically and subscript ``$0$'' marks the value obtained from zero-order trajectory neglecting any Coulomb interaction.
The newly defined relative error is well behaved even for the final vanishing momentum where the momentum is replaced with the total CMT instead.
We show the re-defined error for valid ranges of the ionization phase $u_i$, and the initial transversal momentum  in Fig.~\ref{fig:13:applicability_zero} for our zero-order trajectory method, and in Fig.~\ref{fig:18:applicability_step} for the step-by-step method.

Obviously, the initial momentum $p_{\perp i}$ plays a crucial role.  With decreasing $p_{\perp i}$ the first rescattering has smaller impact parameter and, therefore, induces larger CMT, which will introduce discrepancy to the zero-order trajectory. Thus, our method for analytical estimation of CMT is not applicable for small initial transverse momenta of the ionized electron and for small ionization phases (near the peak of the laser field). The vertical lobes indicate the ionization phases with underlying SR. The error rises there, since the CMT at such recollision is much larger than the CMT at fast recollisions and even a small relative error has a large total contribution. 

Finally, let us note that the white areas arise due to several effects: such as chaotic dynamics, hard recollisions, and the trapping of electrons in Rydberg states. Since such effects are not expected to play a significant role for Coulomb focusing, we are not concerned by them at this point. The role of the errors of the present analytical approach for the description of the final PMD is going to be discussed in the next section.

\subsection{Photoelectron momentum distribution}\label{sec:PMD}

\begin{figure*}
	\centering
  		  		\includegraphics[width=1\textwidth]{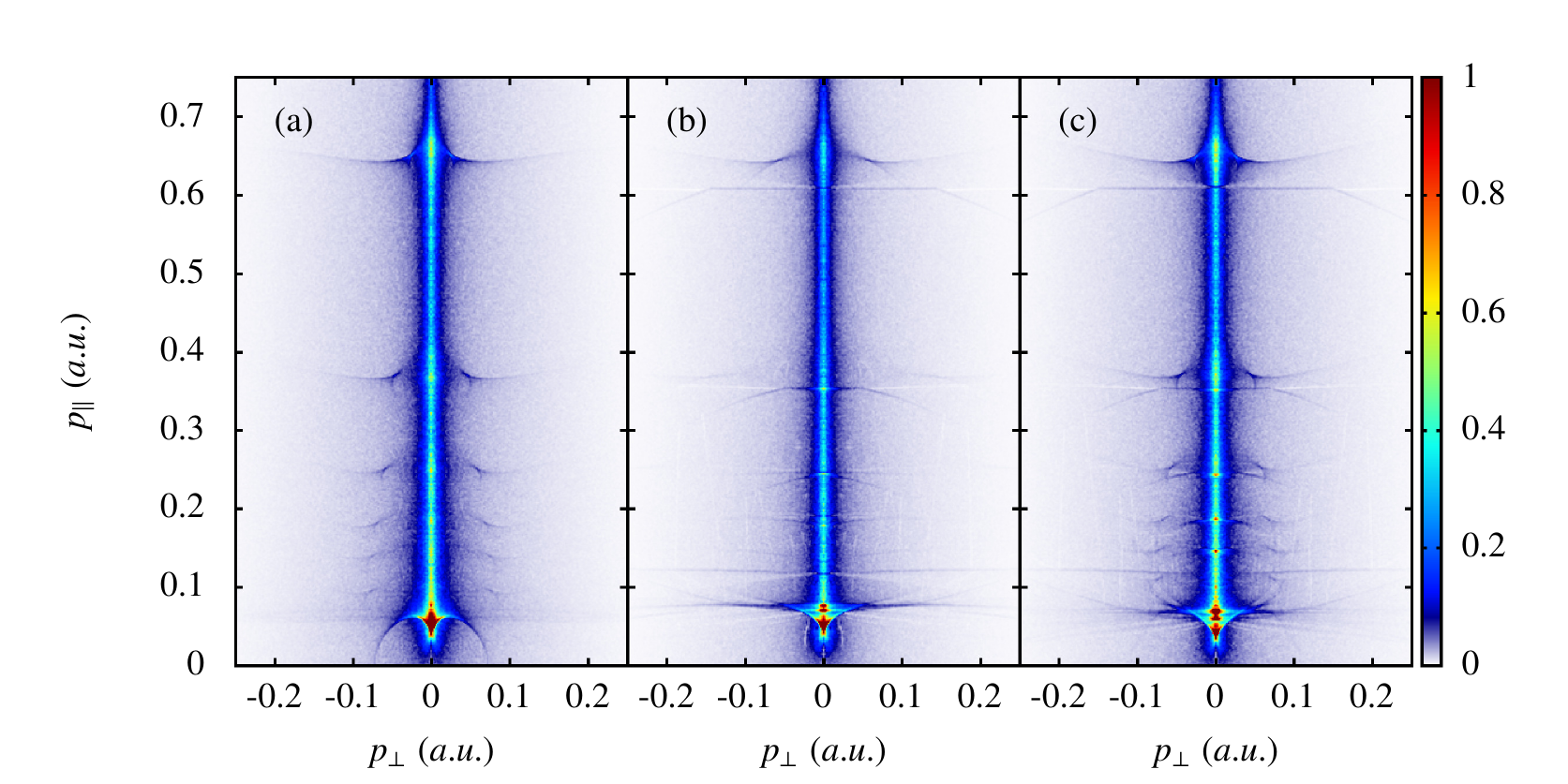}
\caption{PMD: (a) numerical CTMC simulation; (b) via the zero-order trajectory method, and  (c) via the step-by-step method.}
\label{fig:20:PMDs}
\end{figure*}
\begin{figure}
	\centering
  		  		\includegraphics[width=0.9\linewidth]{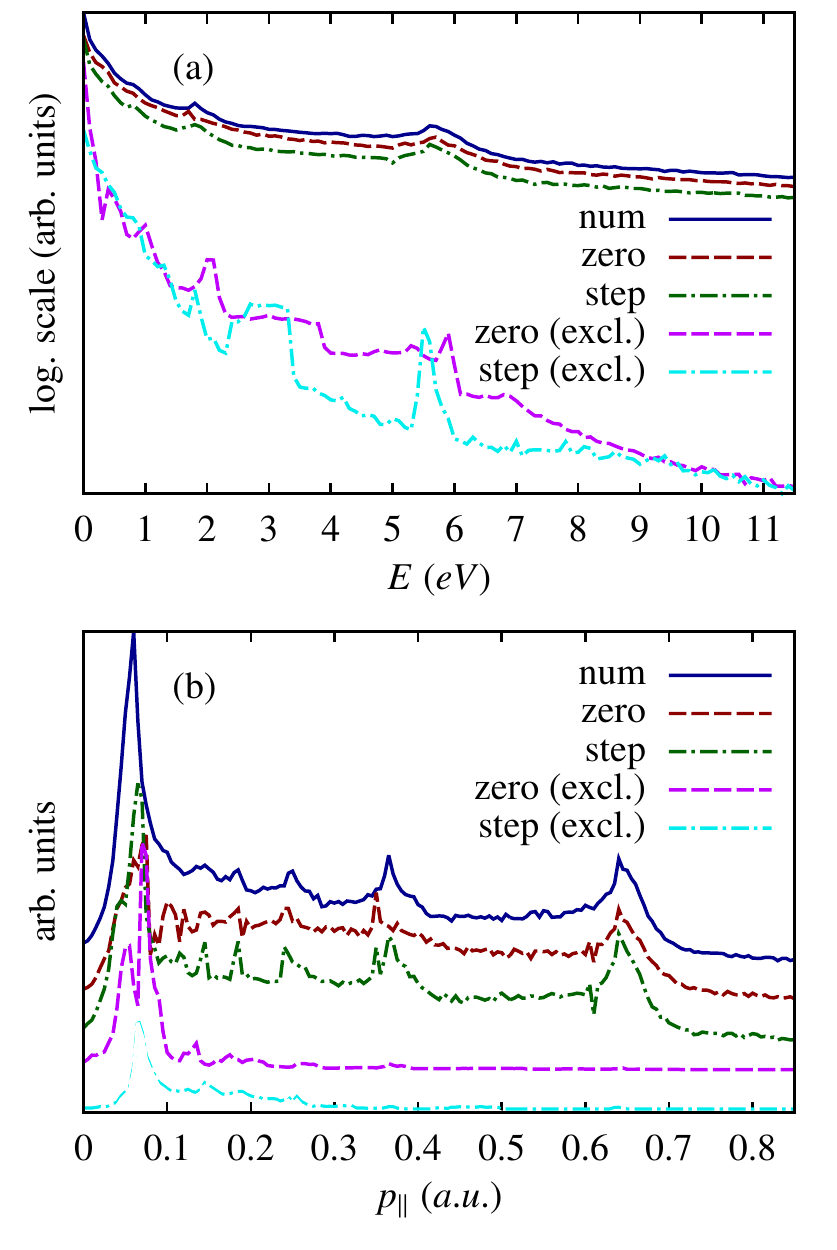}
\caption{ (a) Photoelectron energy distribution; (b) Photoelectron longitudinal momentum distribution: numerical CTMC simulation (solid, blue), via the zero-order trajectory method (dashed, red), via the step-by-step method (dash-dotted, green). 
We show also distributions while restricting trajectories to those with an error greater than 100\% in  at least one directions: via the zero-order trajectory method (dashed, magenta), via the step-by-step method (dash-dotted, cyan). The curves are shifted vertically for visibility while keeping the same shift for the same methods in individual panels.
}
\label{fig:21:En_pxA}
\end{figure}

In the previous section we have seen that our analytical methods allow us to determine the photoelectron asymptotic momentum during the laser driven excursion in the continuum in the field of atomic core. However, since the accuracy is not acceptable in the whole range of the ionization phases or in the initial transverse momenta, a question arises how accurately can the final PMD be described by our methods. In this section we compare results of fully numerical CTMC simulation of the final PMD with those of our analytical methods.
In order to do this, we performed  CTMC simulations in 2D due to the symmetry of the problem in the dipole approximation. 
Every 2D trajectory of the initial transversal momentum $p_{\bot i}$ is weighted with the PPT transverse momentum distribution $ w_{\text{PPT}}(p_{\bot i})$ and with an additional factor of $2\pi p_{\bot i}$. The later accounts for the fully 3D initial phase space whose two transversal dimensions can be mapped onto a single dimension due to the symmetry via $\dd^2{p}_{\bot i}= 2\pi p_{\bot i}\dd p_{\bot i}$.
With the electron asymptotic distribution function $w_{\text{sim}}(p_{\|},p_{\perp})$ provided by the 2D CTMC simulation, one restores the real final 3D PMD: 
\begin{eqnarray}
\frac{\dd^3 f}{\dd^3 {p}}\propto \frac{w_{\text{sim}}(p_{\|},p_{\perp})}{2\pi p_{\bot }},
\label{eq:41:wadk}
\end{eqnarray}
where we have restored the second transversal dimension via the relation $\dd^2{p}_{\bot}= 2\pi p_{\bot}\dd p_{\bot}$ for the final transversal momentum $p_{\bot}$.

We performed three different CTMC simulations with $10^7$ trajectories to determine the PMD at the end of the laser pulse: one fully numerical, second using our zero-order trajectory method, and the last using the step-by-step method. The resulting PMDs are compared in Fig.~\ref{fig:20:PMDs}.

As we can see, both methods reproduce the central vertical cusp. However, the width of the cusp is reproduced by the step-by-step method more correctly. On one hand, the horizontal fringes appear already by the zero-order method, which can be understood as a manifestation of the fine role of the SRs (so called, longitudinal bunching \cite{Kastner_2012}). On the other hand, the step-by-step method seems to reconstruct even the right thickness and location of the fringes and thus concludes as a better investigative tool. Unfortunately, both simulated PMDs exhibit additional horizontal lines (e.g., at $p_x \approx 0.61\;\text{a.u.}$). Such lines can be contributed to an ``artificial" longitudinal bunching effect which arises when one SR is replaced by two FR yielding a slightly greater total CMT. Because of this artifact, an additional horizontal line appears underneath each SR regular fringe which demonstrates then as a twofold line in the PMDs. Luckily, the utility of the results is not jeopardized since the artificial fringe is much weaker than the real effect and can be therefore easily disclosed.

The numerical PMD possesses a half-circle fringe of radius $\sim 0.08\;\text{a.u.}$ (center at the origin) with a prominent peak structure inside. This structure is created by electrons with low transverse momenta near the peak of the laser pulse, for which the error of our analytical methods are large, see white area in Figs.~\ref{fig:13:applicability_zero} and \ref{fig:18:applicability_step}. Although both analytical methods reproduce the peak, they fail to predict the correct structure of it. 
This is due to the fact that these electrons undergo multiple recollisions with large  CMT and never really gain substantial distance from the ion during the whole laser pulse and are strongly influenced by the Coulomb field even when the pulse is long gone. For such behavior, the perturbative recollision picture does not hold and our methods fail.

Since the first observation of the LES was carried out in the photoelectron energy spectra \cite{Blaga_2009}, we show in Fig.~\ref{fig:21:En_pxA} how our analytical methods reproduce the spectra and the longitudinal momentum distribution obtained by the CTMC simulations where we separated the contributions of the electrons creating the PMDs in Fig.~\ref{fig:20:PMDs} and of those which were not taken into account due to significant errors. 
As we see, the excluded contributions are negligible in the energy domain; especially, for non-vanishing energies. However, two features manifesting as sharp peaks can be found at $p_x \approx 0.65\; a.u.$ and at $p_x \approx 0.075\; a.u.$, which can be discerned clearly in in the Fig.~\ref{fig:20:PMDs}.
As we can see, already the zero-order trajectory method captures the positions of the peaks in the energy distribution quite correctly. Nevertheless, the peak at the vanishing longitudinal momentum is misplaced which is corrected by the more precise step-by-step method.

We can conclude that our analytical approach is able to predict correctly many features of PMD, in particular, width of the vertical cusp, the peaks along it, and the position of the horizontal caustic fringes due to the longitudinal bunching. Our approach fails only at very low momenta $p_x\lesssim 0.05$ a.u. While predicting the existence of the lowest momentum peak, none of our methods provides its correct structure. 
The reason is that the trajectories with a large error (white areas in Figs.~{\ref{fig:13:applicability_zero}-\ref{fig:18:applicability_step}) mostly contribute to this prominent peak at low momenta 
which explains the noticeable discrepancy between the numerics and our results in this region. 
On the bright side, the white areas contribute to the momentum peaks at larger energies only negligibly and do not threaten the utility of our approach to CF for the largest part of the PMD, see Fig.~\ref{fig:21:En_pxA}(b).

\section{Non-dipole effects}
\label{sec:non-dipole:07:non-dipole}

In this section we demonstrate how the analytical formulas derived in the previous sections can be employed to gain insight into the process of CF in the non-dipole regime. In particular, we provide an explanation for the observed counterintuitive and energy dependent shift of the CF cusp in PMD. This shift manifests a breakdown of the dipole approximation, because in the dipole approximation the cusp resides at the center of PMD (i.e., $p_\bot=0$). 

\begin{figure}
    \begin{center}
    \includegraphics[width=0.35\textwidth]{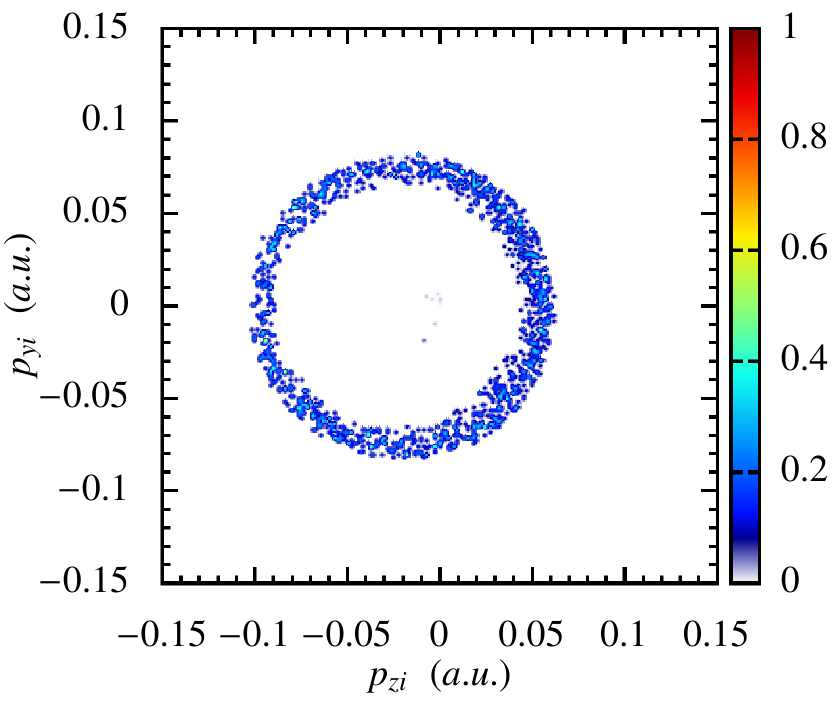} 
   	\caption{ Momentum space of the initial electron distribution at the tunnel exit:  the part which contributes to the final momentum bin at $p_x=0.588$ a.u., and $p_z=-0.0157$ a.u., corresponding to the first soft recollision. Note that the center of the ring is shifted to the negative $p_z$ direction.}
        \label{phase_space}
    \end{center}
  \end{figure}

Our aim is to investigate 
the origin of the shift and to explain its nontrivial shape in the non-dipole regime. The central cusp is shifted along the laser propagation direction by a value varying with respect to the asymptotic longitudinal momentum which can be seen in our CTMC simulation of PMD presented in Fig.~\ref{fig:20:PMD:non-dipole:noKepler}. 

For large longitudinal momenta the shift is positive, which is expected, because of the positive drift momentum of the electron in the laser field along the propagation direction. Meanwhile, for lower longitudinal momenta the shift becomes negative, but tends again to zero at very low longitudinal momenta. Such complex momentum dependence is intriguing and demands further investigation. 


It is known that the electrons ending on the central cusp originate on a circle in the initial transverse momentum phase space at the ionization tunnel exit \cite{Maurer_2017}. Moreover, the center of this circle gets shifted against the propagation direction in the non-dipole regime as a compensation for the magnetically induced momentum drift as shown in Fig.~\ref{phase_space}. The radius of this ring in the initial momentum space is an indicator of CF because it 
equals to the total transverse R-CMT for the cusp electrons. 

Therefore, let us follow an electron which originates at the tunnel exit with $p_{zi}\approx 0$ and $p_{yi}\neq 0$, and ends up at the cusp asymptotically. The final $z$-component of the photoelectron momentum is according to Eq.~(\ref{eq:pz_drift_recollpoint}):
\begin{eqnarray}
p_{zf}\approx  p_{1z}+\frac{A^2(u_i)}{2c}-\frac{p_{xi}A(u_i)}{c},\label{pzf}
\end{eqnarray}
where $p_{1z}$ is the R-CMT for the discussed electron ionized at the laser phase $u_i$, the laser wave propagate along the $z$-axis, and is polarized along the $x$-axis.  While the electron dynamics along the $z$-axis is modified only by the laser magnetic field in this setup, in the $y$-direction it is similar to the dipole case and the final $y$-component of the electron momentum is vanishing:
\begin{eqnarray}
p_{yf}=p_{yi}+  p_{2y,\text{in}}+p_{1y}\approx 0,
\label{pyf}
\end{eqnarray}
where $p_{yi}$, $p_{2y,\text{in}}$ and $p_{1y}$ are the $y$-components of the electron initial momentum, of I-RMT, and of the total R-CMT, respectively.
Assuming that the electron undergoes $N$ rescatterings, Eq.~(\ref{pyf}) reads
\begin{equation}
p_{yi} + p_{2y,\text{in}} + \sum\limits_{n=1}^{N} p^{(n)}_{1 y} \approx 0,
\label{eq:nondip:bend_aver_cond}
\end{equation}
where $p^{(n)}_{1 y} $ is the R-CMT at the $n^{th}$ recollision.

Further, we exploit the possibility to estimate the individual R-CMT as a product of the acting force during the recollision and the time of the recollision $\tau_n$ as shown in Sec.~\ref{sec:dipole:04:solving_first_order}: 
\begin{equation}
 p^{(n)}_{1 y}\approx -Z\frac{y_{n}}{r_{n}^{3}}\tau_{n}.
\label{eq:app:non-dipole:cond_on_py}
\end{equation}
Later we can apply the derived analytical formulas Eqs.~(\ref{eq:14:corr_peak_preciser_generealized_y})-(\ref{eq:14:corr_peak_preciser_generealized_trans1}), (\ref{eq:11:pz_corr_plateau_preciser_TP_asy_y})-(\ref{eq:11:pz_corr_plateau_preciser_TP_asy_z}) and (\ref{eq:11:py_corr_plateau_preciser_turnpoins})-(\ref{eq:11:pz_corr_plateau_preciser_turnpoins}) to estimate recollision time $\tau_{n}$ precisely.

\begin{figure}
	\centering
  		\includegraphics[width=0.5\textwidth]{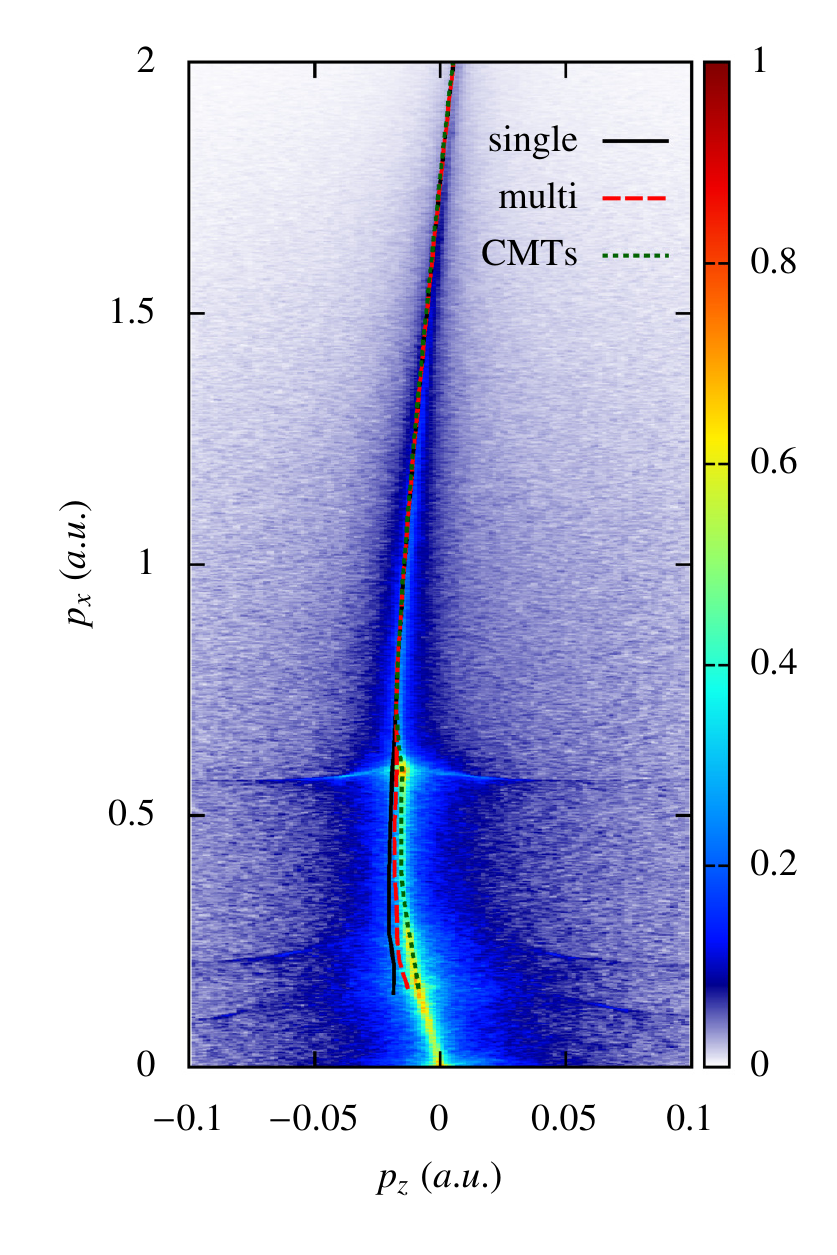}
\caption{ The manifestation of the breakdown of the dipole approximation in PMD: CTMC simulation for hydrogen in intense laser pulse of linear polarization  
$E_0 = 0.0407$, $\omega = 0.012$ and the laser pulse duration $15.7/\omega$. The peak of the cusp was estimated via the first term in Eq.~(\ref{pzf2}), $p_{zf}=-\overline{T}_{zd}^{(1)}$ (black, solid), further via the full Eq.~(\ref{pzf2}) (red, dashed), and finally calculated as a sum of  I-CMT and R-CMT at each recollision via analytical formulas of Sec.~\ref{sec:dipole:04:solving_first_order} (green, dotted).}
\label{fig:20:PMD:non-dipole:noKepler}
\end{figure}
\begin{figure}	\centering  		\includegraphics[width=0.5\textwidth]{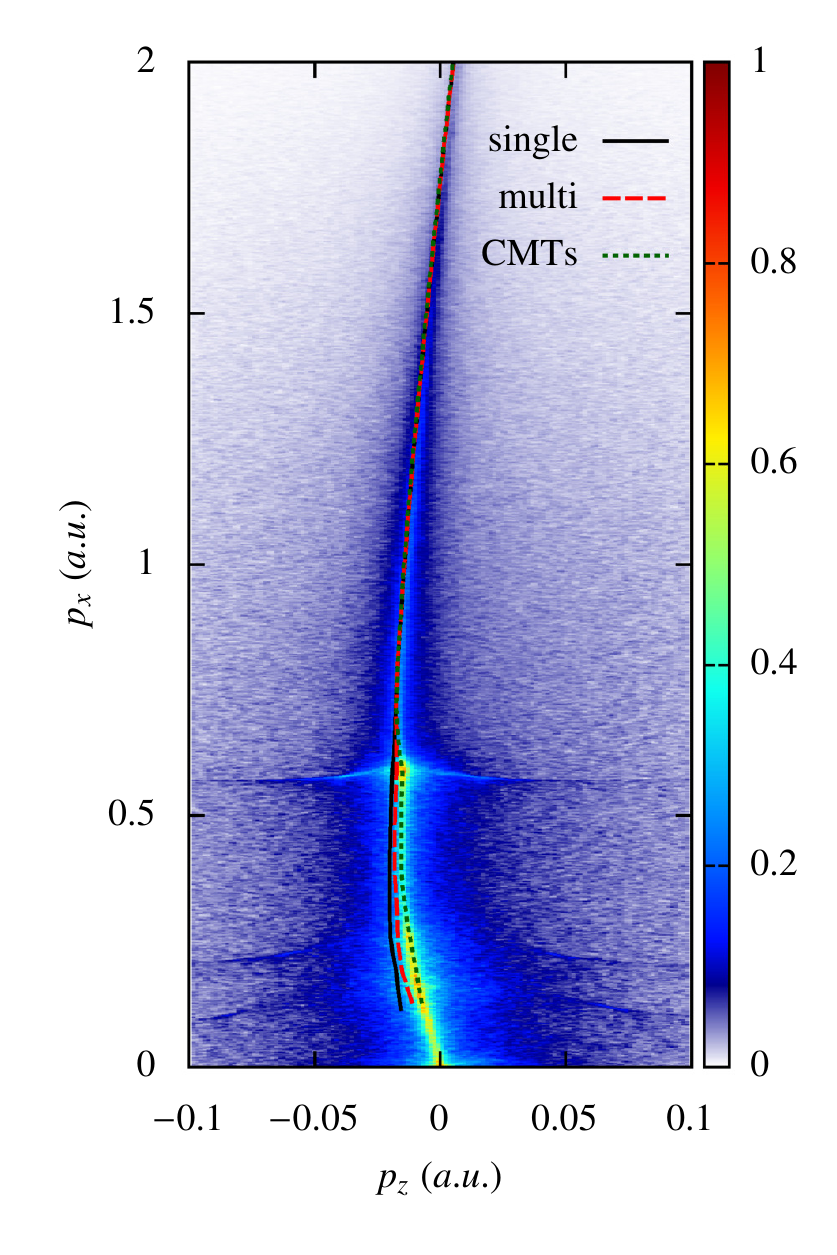}\caption{The same as Fig.~\ref{fig:20:PMD:non-dipole:noKepler} but in the estimation of the position of the cusp, the Coulomb field effect after switching off the laser pulse has been taken into account additionally. }
\label{fig:20:PMD:non-dipole:Kepler}
\end{figure}
The change to the final momentum $ p_{1z}$ due to the Coulomb interaction is estimated as the sum of R-CMTs  yielding 
\begin{equation}
 p_{1z} = \sum\limits_{n=1}^{N} p_{1 z}^{(n)} = - Z\sum\limits_{n=1}^{N} \frac{z_{n}}{r_{n}^{3}}\tau_{n}.
\label{eq:app:non-dipole:delta_pz_f_definition}
\end{equation}
As the recollision time $\tau_{n}$ and the recollision distance $r_{n}$ are the same in Eqs.~(\ref{eq:nondip:bend_aver_cond}) and (\ref{eq:app:non-dipole:delta_pz_f_definition}), we may use the former to simplify the expression for $p_{1z}$. For the latter we need to estimate the recollision coordinates $y_n$ and $z_n$. We define them step-wise from recollision to recollision as 
\begin{eqnarray}
y_1 &=& \left(p_{yi} + p_{2y,\text{in}}\right)\left(t_1-t_0\right), \nonumber\\
y_2 &=& \left(p_{yi} + p_{2y,\text{in}}\right)\left(t_2-t_0\right) - Z\frac{y_1}{r_1^3}\tau_{n}\left(t_2-t_1\right), \nonumber\\
&\vdots & \nonumber\\
y_N &=& \left(p_{yi} + p_{2y,\text{in}}\right)\left(t_N-t_0\right) - Z\sum\limits_{n=1}^{N-1}\frac{y_n}{r_n^3}\tau_{n} \left(t_{N} - t_{n}\right),\label{eq:app:non-dipole:y_N_general}
\end{eqnarray}
where $t_0$ is the ionization time, and $t_n$ is the $n^{th}$ recollision time for $n\geq 1$.
The upper formulas can be used to express  Eq.~(\ref{eq:nondip:bend_aver_cond}) as follows
\begin{eqnarray}
p_{yi} + p_{2y,\text{in}} &=& Z \sum\limits_{n=1}^{N}\frac{\tau_{n}}{r_n^3}\left(p_{yi}+p_{2y,\text{in}}\right)\left(t_n-t_0\right)\nonumber\\
 &&
 - Z^2\sum\limits_{n=1}^{N}\frac{\tau_{n}}{r_n^3}\sum_{k=1}^{n-1}\frac{y_k}{r_k^3}\tau_{k} \left(t_{n} - t_{k}\right),
\label{eq:app:non-dipole:cond_on_py_subs}
\end{eqnarray}
where the first term corresponds to first order corrections to the zero-order trajectory and the second iterative term to the next order corrections. 
When neglecting higher-order correction terms than the second-order proportional to $\sim \left(\tau_{n}/r_n^3\right)^{2}$, we arrive after rearrangement at 
\begin{eqnarray}
Z \sum\limits_{n=1}^{N}\frac{\tau_1}{r_n^3} \left(t_1 - t_0 \right) \approx 1 + Z^2\sum\limits_{k<n}^{N}\frac{\tau_n\tau_k}{r_n^3r_k^3}\left(t_{n}-t_{0}\right)\left(t_{n}-t_{k}\right). 
\label{eq:app:non-dipole:cond_on_py_rearranged}
\end{eqnarray}
The rescattering coordinate along the laser propagation direction $z_r$ depends on the magnetically induced drift momentum of the electron ($p_{zi}=0$ is chosen):
\begin{eqnarray}
z_1 &=& \overline{p}^{(1)}_{zd} \left(t_1 - t_0\right), \nonumber\\
z_2 &=& \overline{p}^{(2)}_{zd} \left(t_2 - t_0\right) - Z\frac{z_1}{r_1^3}\tau_{n}\left(t_2-t_1\right), \nonumber\\
&\vdots & \nonumber\\
z_N &=& \overline{p}^{(N)}_{zd} \left(t_N - t_0\right) -Z \sum\limits_{n=1}^{N-1}\frac{z_n}{r_n^3}\tau_{n} \left(t_{N} - t_{n}\right),
\label{eq:app:non-dipole:z_N_general}
\end{eqnarray}
where we used the magnetic drift from Eq.~(\ref{eq:pz_drift_recollpoint}) 
to define the averaged drift momentum as 
\begin{equation}
\overline{p}^{(n)}_{zd} \equiv \frac{1}{t_{n}-t_0}\int\limits_{t_{0}}^{t_{n}} 
\left\{\frac{p_{x i}}{c}\left[A_x( t)-A_x( t_0)\right]+\frac{1}{2c}\left[A_x( t)-A_x( t_0)\right]^2 \right\}\dd t.
\label{eq:app:non-dipole:pzdrift_aver_def}
\end{equation}
Although in the tunneling regime $p_{x i} = 0$,  we incorporate the I-CMT of Eq.~(\ref{correctedICMTxit}) into $p_{x i}$ and set $p_{x i} = p_{2 \parallel, \text{in}}(u_i, p_{y i})$.
When we substitute Eq.~(\ref{eq:app:non-dipole:z_N_general}) into the Eq.~(\ref{eq:app:non-dipole:delta_pz_f_definition}), we obtain
\begin{eqnarray}
p_{1z} &=& -Z\sum_{n=1}^{N}\overline{p}_{zd}^{(n)} \frac{\tau_{n}}{r_n^3} \left(t_n-t_0\right) \nonumber\\
& & + Z^2\sum_{n=1}^{N}\frac{\tau_{n}}{r_n^3}
\sum_{k=1}^{n-1}\frac{z_{k}}{r_k^3}\tau_{k} 
\left(t_{n} - t_{k}\right),
\label{eq:app:non-dipole:delta_pz_f_substit}
\end{eqnarray}
where the first term again corresponds to first-order correction to unperturbed trajectory and the second term to higher-order corrections. When neglecting the higher-order contributions $\sim \left(Z\tau_{n}/r_n^3\right)^{3}$ and employing the relationship of Eq.~(\ref{eq:app:non-dipole:cond_on_py_rearranged}), we can write 
\begin{eqnarray}
 p_{1z} \approx -\overline{p}_{zd}^{(1)} + Z\sum\limits_{n=1}^{N}\left(\overline{p}_{zd}^{(1)}-\overline{p}_{zd}^{(n)}\right)\frac{\tau_{n}\left(t_{n}-t_{0}\right)} {r_n^3}.
\nonumber\\
\label{eq:app:non-dipole:delta_pz_f_final}
\end{eqnarray}
The first term here describes R-CMT at the first recollision and the other terms arise only due to the multiple recollisions.

In the case of a single recollision, the lengthy derivation above becomes very transparent:
\begin{equation}
 p_{1z} =   -  Z \frac{z_{1}}{r_{1}^{3}}\tau_{1}= -\frac{z_1}{y_1}p_{1y}=-\frac{z_1}{t_1-t_0}=-\overline{p}_{zd}^{(1)},
\label{sigle_recollision}
\end{equation}
showing that in this case the z-component of R-CMT equals the averaged (between the ionization and the rescattering time) drift momentum in the laser propagation direction.

One can give another intuitive perspective to  Eq.~(\ref{sigle_recollision}) of the single recollision case while discussing the non-dipole dynamics of the electron in the cusp, which has an initial momentum only along the laser propagation direction $\textbf{p}_{i}=(0,0,p_{iz})$, and comparing it with the dipole case. The electron dynamics in the non-dipole and in the dipole case will be similar, if the recollision impact parameters are the same $z_r=z_r^{(0)}$, where the parameters of the dipole case are indicated with the upper index $^{(0)}$. 
The rescattering coordinate in the non-dipole case is $z_r=(p_{zi}+\overline{p}_{zd})(t_1-t_0)$, while in the dipole case simply $z_r^{(0)}=p_{zi}^{(0)} (t_1-t_0)$. Thus, the electron dynamics will be similar in both cases if  $p_{zi}=p_{zi}^{(0)} -\overline{p}_{zd}$, i.e., when the electron has an additional initial momentum opposite to the laser  magnetically induced drift. The similar dynamics means, in particular, the same R-CMT: $p_{1z}\approx p_{1z}^{(0)}$. Because in the non-dipole case the cusp is at vanishing momentum, i.e., $ p_{ zi}^{(0)}+p_{1z}^{(0)}\approx 0$, we arrive at relation $p_{1z} = -p_{zi} - \overline{p}_{zd}$, corresponding to Eq.~(\ref{sigle_recollision}) with $p_{zi}\approx 0$. The comparison links the averaged drift momentum before the first rescattering directly to the asymptotic momentum of the cups electrons which can now be estimated via Eq.~(\ref{pzf}) as
\begin{equation}
p_{zf}=p_{zi}+p_{1z}+\frac{A^2(u_i)}{2c} \approx -\overline{p}_{zd}+\frac{A^2(u_i)}{2c},
\label{eq:app:non-dipole:pz_f_final_single_RP}
\end{equation} 
where an intriguing interplay of the R-CMT and of the magnetic drift contributions given by their opposite signs is revealed.

\begin{figure}[b]
	\centering
  		\includegraphics[width=0.4\textwidth]{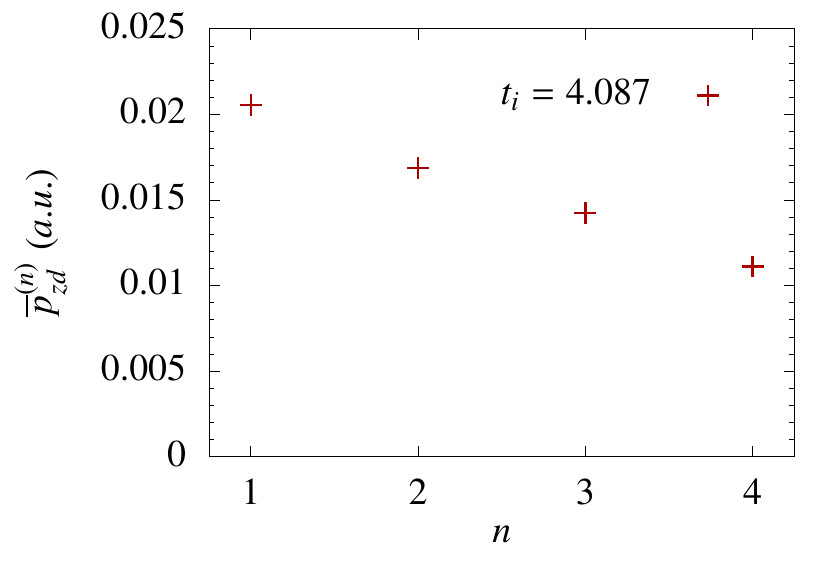}		
\caption{The average magnetically induced drift momentum $\overline{p}_{zd}^{(n)}$ obtained till the $n^{\text{th}}$ rescattering for the ionization time $t_i=4.087$, corresponding to $p_x \approx 0.3$ and hence to multiple rescattering case.
}
\label{fig:20:PMD:non-dipole:R-CMT}
\end{figure}

In the general case of multiple recollisions we derive the asymptotic momentum using Eq.~(\ref{eq:app:non-dipole:delta_pz_f_final}) as
\begin{equation}
p_{zf} \approx \frac{A^2(u_i)}{2c}-\frac{p_{xi}A(u_i)}{c} -\overline{p}_{zd}^{(1)} + Z \sum\limits_{n=1}^{N}\left(\overline{p}_{zd}^{(1)}-\overline{p}_{zd}^{(n)}\right)\frac{\tau_{n}\left(t_{n}-t_{0}\right)}{r_n^3},
\label{eq:app:non-dipole:pz_f_final}
\end{equation}
where we sum over $N$-recollisions.
The first three terms in this equation can be combined yielding
\begin{eqnarray}
p_{zf} \approx -\overline{T}_{zd}^{(1)}+Z \sum\limits_{n=1}^{N}\left(\overline{p}_{zd}^{(1)}-\overline{p}_{zd}^{(n)}\right)\frac{\tau_{n}\left(t_{n}-t_{0}\right)}{r_n^3},\label{pzf2}
\end{eqnarray}
where
\begin{eqnarray}
\overline{T}_{zd}^{(1)}\equiv \frac{1}{u_1-u_0}\int_{u_0}^{u_1}\left[ \frac{p_{xf}A(u)}{c}+\frac{A^2(u)}{2c}\right]\dd u,
\end{eqnarray}
with $u_0 = u_i$ and asymptotic longitudinal momentum $p_{xf}\equiv  - A(u_i) + p_{xi}$. We would also like to stress the not evident coincidence of
the term $-\overline{T}_{zd}^{(1)}$ with the single rescattering of case of Eq.~(\ref{eq:app:non-dipole:pz_f_final_single_RP}).

We demonstrate our results in Fig.~\ref{fig:20:PMD:non-dipole:noKepler}, where we compare them to the CTMC simulation. 
Let us note at this point that all rescattering parameters used thorough this section were obtained from the numerical trajectories.

The first term, $-\overline{T}_{zd}^{(1)}$,  in Eq.~(\ref{pzf2}) describes the main contribution to the negative shift of the cusp which is due to the first recollision. It coincides with the result of Ref.~\cite{Tao_2017} and describes very well the peak of the cusp for large longitudinal momenta ($p_x \gtrsim 0.52$) when only a single rescattering exists. However, a relatively large discrepancy appears at lower momenta, when the negative shift of the cusp begins to decrease, tending to zero at very low energies. The horizontal fringes in PMD correspond to the slow recollision condition. When one crosses the horizontal line towards lower longitudinal momenta, the number of rescatterings increases by one. 


When there are more than one rescattering, which is the case for $p_x \lesssim 0.52$ in Figs.~\ref{fig:20:PMD:non-dipole:noKepler} and \ref{fig:20:PMD:non-dipole:Kepler}, the non-dipole dynamics is no more similar to the dipole case. In fact, by a single parameter of the initial transverse momentum, one cannot fix the impact parameters of multiple scattering events to be the same as in the dipole case.  The role of the multiple recsatterings is given by the second term in Eq.~(\ref{pzf2}) expressed as a sum. It is positive, see an example of the estimation for certain trajectory in Fig.~\ref{fig:20:PMD:non-dipole:R-CMT}, and yields the decrease of the negative shift of the cusp.


\begin{figure}
	\centering
  		\includegraphics[width=0.4\textwidth]{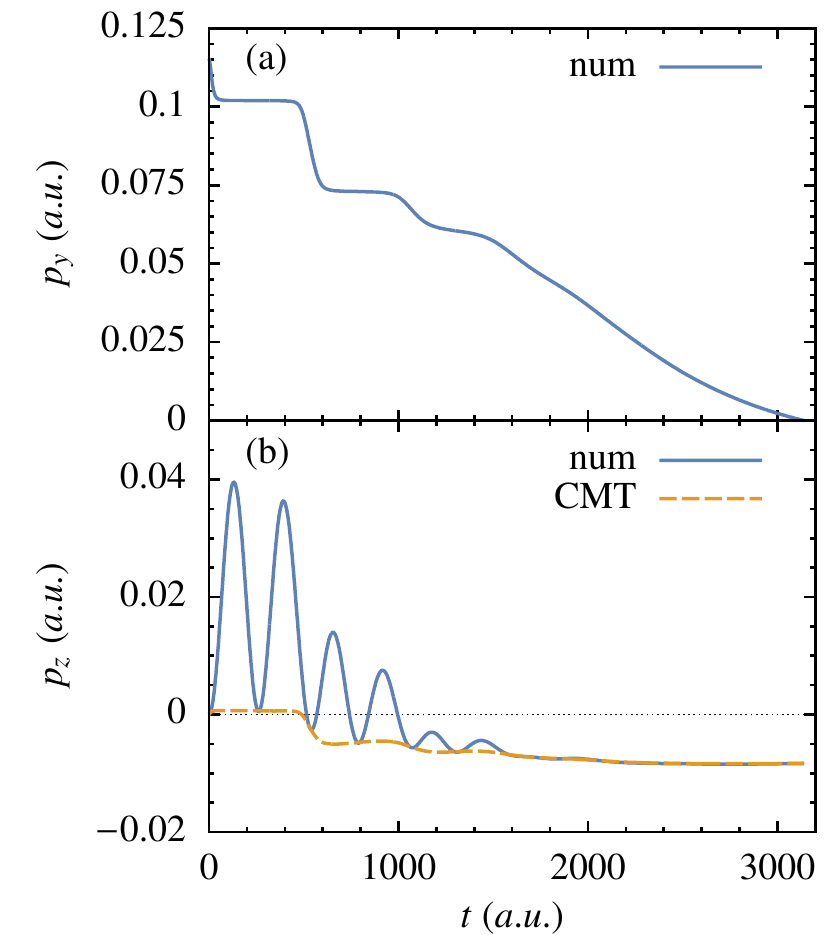}
\caption{The evolution of the compensated momentum of the electron $\textbf{p}(t)+\textbf{A}(t)$, which indicates the history of the Coulomb momentum transfer. The example shows the trajectory which ends up at the PMD cusp,   tunneled at $t_{i} = -1.25$ with $p_{yi} = 0.116$ and $p_{zi} = 0$. Panel (a), the $p_y$ momentum component changes significantly at the end of the laser pulse ($t\approx 2200$)
Panel (b) shows  negative R-CMT of the $p_z$ at the end of the laser pulse where we isolated the contribution of the Coulomb interaction and plotted it by dashed yellow line.
}
\label{fig:20:PMD:non-dipole:R_CMT-numerics}
\end{figure}

Whereas the positive offset of the bend can be attributed to the magnetic drift $p_{zd} = A^2(u_{i})/2c$, the negative offset can be seen as a result of a counterbalance between the Coulomb momentum transfer $p_{1z} $ and magnetically induced drift $p_{zd}(u,u_i) $. Due to the magnetically induced drift the electron moves in the positive $z$-coordinate direction and gains a negative Coulomb momentum transfer in $z$-direction at the rescattering; the larger the drift (which is the case for the smaller longitudinal momentum), the larger the R-CMT becomes in $z$-direction.
Such picture holds for relatively large longitudinal momenta with a single recollision.
Nevertheless, the simulations have shown that the offset of the cusp at lower longitudinal momenta bends again towards laser propagation direction which is related to the increasing number of recollisions. The effective drift in this case is decreasing as $\overline{p}^{(n)}_{zd}< \overline{p}^{(1)}_{zd}$, which consequently decreases the total CMT in $z$-direction. 

There is a second reason for the PMD cusp to shift towards vanishing momentum at very low energies. The low energy electrons are still rather close to the parent ion when the interaction with the laser pulse is over. 
At this point and later on, the focusing property of the Coulomb field manifests and pulls these electrons towards the ion, decreasing further the transverse momentum, see an example in Fig.~\ref{fig:20:PMD:non-dipole:R_CMT-numerics}. As we can see in Fig.~\ref{fig:20:PMD:non-dipole:R_CMT-numerics}(a), the $p_y$ momentum component changes significantly even at the end of the laser pulse, where the recollision picture does not hold anymore, and the condition given in Eq.~(\ref{eq:nondip:bend_aver_cond}) is not applicable. Part (b) shows nontrivial negative CMT for the $p_z$ at the tail of the laser pulse which is not incorporated in our simplified discussion in this section. 
Since the breakdown of the recollision picture poses a nontrivial problem to account for, we included at least the additional role of the Coulomb potential after the end of the laser pulse in Fig.~\ref{fig:20:PMD:non-dipole:Kepler}. 
In the figure we estimated the position of the cup by the asymptotic Kepler formula using the position from the numerical trajectory at the end of the laser pulse and the momenta from the Eqs.~(\ref{eq:app:non-dipole:pz_f_final_single_RP}), (\ref{pzf2}).
As we can see, only the low-energetic trajectories near the origin are noticeably influenced yielding a better agreement with the simulation in this region.

\section{Conclusion}
\label{sec:dipole:08:discussion}

We have developed an analytical  model for quantitative description of  CF effects in laser induced strong field ionization. 
Under the assumption that the Coulomb field effect is a perturbation for the near recollision laser driven trajectory, we have derived past-free analytical formulas for the Coulomb momentum transfer at recollisions which depend on the local recollision coordinate and momentum. Moreover, for an effective treatment of Coulomb momentum transfer we classify the recollisions into two types: slow- and fast-recollisions. The obtained formulas for the momentum transfer at slow recollisions, Eqs.~(\ref{eq:14:corr_peak_preciser_generealized_long1})-(\ref{eq:14:corr_peak_preciser_generealized_trans1}), and for fast recollisions Eqs.~(\ref{eq:12:corr_plateau_simplest_generealized_long1})-(\ref{eq:12:corr_plateau_simplest_generealized_trans1_z}) 
are applied even in the case when the Coulomb field is not a perturbation  to the global trajectory. In this case the recollision parameters can be derived either by the step-by-step method, or via exact numerical trajectory. The non-dipole effects are shown to be negligible during the brief time of the recollision, however they are indirectly incorporated in the theory via the recollision coordinate and momentum. Within the same model we derived  essential higher-order corrections to the known expressions for the initial Coulomb momentum transfer at the tunnel exit \cite{Shvetsov-Shilovski_2009}. And, while applying perturbation theory for the Coulomb field globally with respect to the laser driven trajectory, the Coulomb momentum transfer has been expressed via the ionization phase and the initial transverse momentum at the tunnel exit for very special but in the literature widely discussed classes of recollisions exposing scaling dependencies of the laser field parameters.

The derived analytical formulas for the Coulomb momentum transfer, employed along with numerical simulations, can help to gain insight into the  detailed features of the CF effect in different laser field setups. In particular, they allow estimation of the role of each particular rescattering event, which is hidden in the fully numerical CTMC simulation, but essentially helps to develop an intuitive picture for CF. In this context, we have proven by our analytical approach (see Fig.~\ref{fig:14:total_mom_corr_Piecewise_arbitrary_E_0_04}) that single rescattering is not sufficient to quantitatively describe CF in mid-infrared laser fields and the contribution of high-order rescatterings  should not be neglected.

We have analyzed the accuracy of our analytical approach in the dipole approximation case, estimating
the total Coulomb momentum transfer 
during multiple recollisions and deriving the final PMD by two various methods. Besides the simplest zero-order laser driven trajectory method, we put forward also a step-by-step method, when after each recollision the electron trajectory is corrected, appropriately revising the electron momentum by the Coulomb momentum transfer at the recollision.  We show that both methods satisfactory describe  the asymptotic PMD in a large range of momentum space, with the step-by-step method describing more closely the fine features of PMD. 
However, the accuracy of our approach fails at very low photoelectron energies, where both methods correctly predict existence of low-energy peak, but do not deliver its correct structure.


Finally, we have employed the derived analytical formulas for R-CMT to gain insight into the nontrivial features of the PMD in the non-dipole regime. The peak of the cusp in the PMD in the non-dipole regime is positive at large longitudinal momenta, but becomes negative at intermediate values, and moves again towards laser propagation direction for even lower rates. We have explained this counterintuitive behavior of the peak of the cusp by relating it to the laser magnetically induced average drift. The increase of the negative shift of the cusp at decreasing longitudinal momentum is explained by the increase of the laser magnetically induced average drift during the single recollision. In the meantime, the damping of the negative shift of the cusp  at very low longitudinal momenta is explained due to the suppression of the effective drift during multiple rescatterings. Additionally, the asymptotic effect of the Coulomb field after switching off the interaction with the laser pulse is shown to contribute to the shifting of the cusp towards vanishing transverse momenta.

\appendix

\section{Calculation of the integral}
\label{sec:app:special_integral}
While computing the general R-CMT at SR for Eqs.~(\ref{eq:14:corr_peak_preciser_generealized_long1})-(\ref{eq:14:corr_peak_preciser_generealized_trans1}), we  applied the tabular integral \cite{GradshteynRyzhik}
\begin{eqnarray}
\int\limits_{0}^{\infty}\frac{ x^{\mu-1} \dd x}{\left(1+2x\cos{t} + x^2\right)^{\nu}} = 
\left(\frac{2}{|\sin{t}|}\right)^{\nu-1/2} \Gamma\left(\nu + \frac{1}{2}\right)\nonumber\\
\times B(\mu,2\nu-\mu)P_{\mu-\nu-1/2}^{1/2-\nu}(\cos{t}),
\label{eq:app:01:10-integral_legendre}
\end{eqnarray}
where $\Gamma(x)$ stands for the Gamma function, $B(x,y)$ for the Beta function, $P_{\nu}^{\mu}(x)$ for the Legendre function of the first kind and $-\pi < t < \pi$, $0 < \Re(\mu) < \Re (2\nu)$.

The Legendre function of the first kind can be expressed for real $x\in \left[-1,1\right]$ as 
\begin{eqnarray}
P_{\nu}^{\,\mu}(x) & = & \frac{1}{\Gamma(1-\mu)}\left(\frac{1+x}{1-x}\right)^{\frac{\mu}{2}}\,
_2F_1\left(-\nu, \nu + 1; 1 - \mu; \frac{1-x}{2}\right),\nonumber\\
\end{eqnarray}
via the hypergeometric function $_2F_1(a,b;c;z)$   \cite{GradshteynRyzhik}, which gives
\begin{eqnarray}
P_{-\frac{3}{2}}^{\,-1}(0) = & \,_2F_1\left(\frac{3}{2}, -\frac{1}{2}; 2; \frac{1}{2}\right) & = 0.786894,
\label{eq:app:P_three_halfs}
\\
P_{-\frac{1}{2}}^{\,-1}(0) = &  \,_2F_1\left(\frac{1}{2}, \frac{1}{2}; 2; \frac{1}{2}\right) & = 1.07871.
\label{eq:app:P_one_halfs}
\end{eqnarray}

\section{The accuracy of the analytical formulas}
 
We have derived several approximative formulas for R-CMT at SR and FR in the section \ref{sec:dipole:04:solving_first_order} and higher order correction to the I-CMT in the section \ref{sec:dipole:05:intial_mom_transfer}. Since several assumptions were made, we have to address the question of agreement between our formulas the exact numerical values in this section. We assume a strong field tunnel ionization of a neutral hydrogen atom and hence we use $Z=1$ a.u. and $I_p = 0.5$ a.u. thorough this section.

\subsection{Recollisions}\label{sec:app:accuracy-R}

The accuracy of the expressions  Eqs.~(\ref{eq:12_px_corr_peak_preciser_alpha_infinity}), (\ref{eq:12_pz_corr_peak_preciser_alpha_infinity}) for SF are analysed in Fig.~\ref{fig:02:peak_comparison_corrections-E_0_04}, where a comparisons with the exact numerical results and with the simple man formulas from Eqs.~(\ref{p1perppeak}),(\ref{p1longpeak}) are given. The longitudinal and also the transverse R-CMT are described well, both with an accuracy $\lesssim20\%$ up to the $10^{\text{th}}$ recollision when R-CMT is decreased by an order of magnitude with respect to the first recollsion. 
From the analysis one can even deduce that the R-CMT estimation is more accurate when ${\cal P}_2=1$ in the corresponding equations.

The analytical results for FR Eqs. (\ref{eq:12_px_corr_peak_preciser_turnpoints}), (\ref{eq:12_pz_corr_peak_preciser_turnpoints}) are compared with the exact numerical calculations and even with the simpler formulas Eqs.~(\ref{eq:11:px_corr_plateau_preciser}), (\ref{eq:11:pz_corr_plateau_preciser}) in Fig. \ref{fig:03:plateau_comparison_corrections-E_0_04}.  As we can see,  simple Eqs. (\ref{eq:11:px_corr_plateau_preciser}), (\ref{eq:11:pz_corr_plateau_preciser}) are more accurate for several first rescatterings (up to the $3^{\text{th}}$ for the longitudinal and $5^{\text{th}}$ for transversal) where the recollision picture holds.  Meanwhile, the formulas (\ref{eq:12_px_corr_peak_preciser_turnpoints}), (\ref{eq:12_pz_corr_peak_preciser_turnpoints}) provide better accuracy for high-order recollisions  where the rescattering picture breaks down. 
\begin{figure}
		\centering
  		  		\includegraphics[width=1\linewidth]{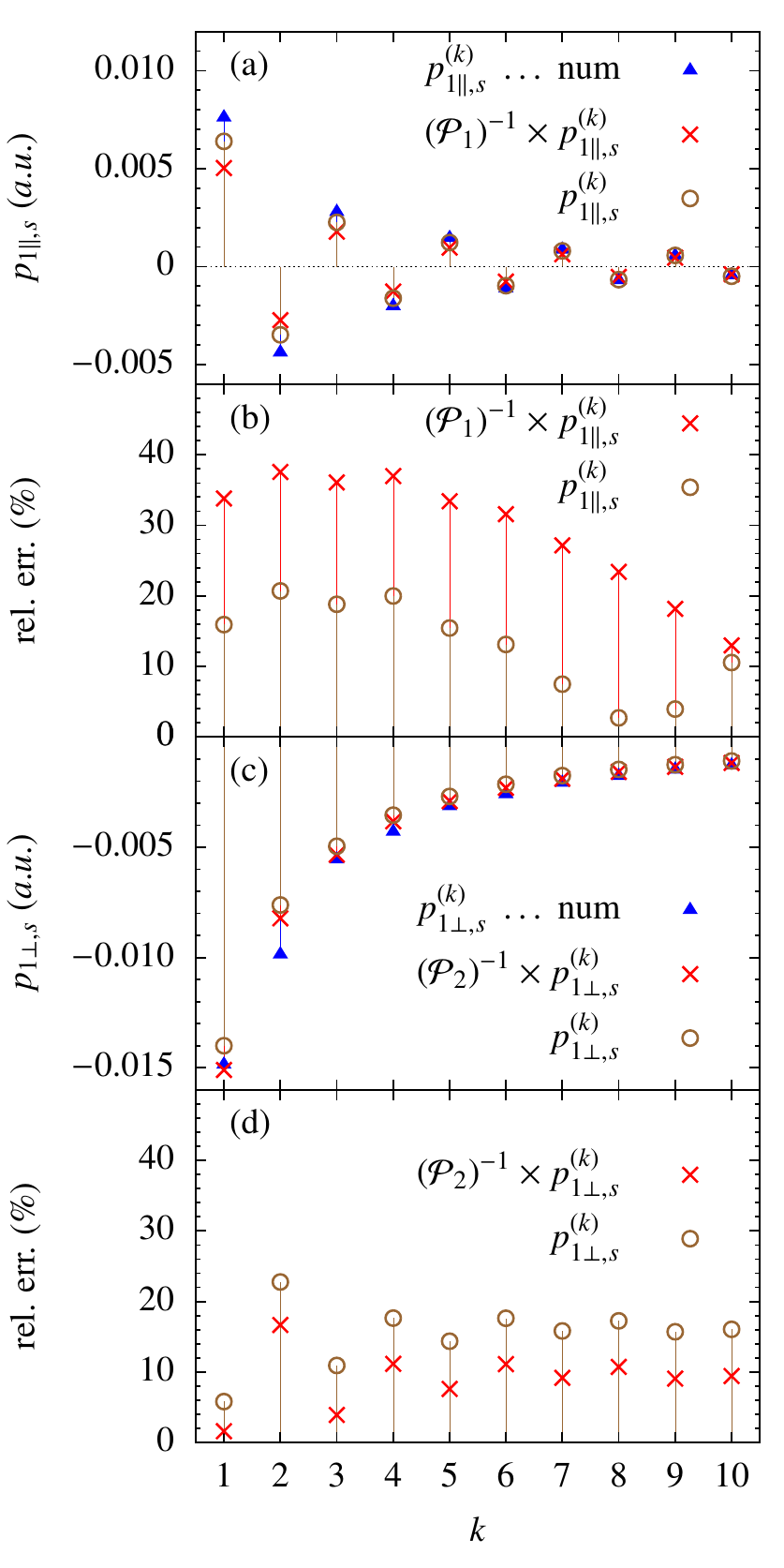}
  		 \caption{Graphical analysis of accuracy of R-CMT estimations at the $k^{\text{th}}$ SR: (brown circles) estimations via Eqs.~(\ref{eq:12_px_corr_peak_preciser_alpha_infinity}), (\ref{eq:12_pz_corr_peak_preciser_alpha_infinity}); (red crosses) simplified estimations via  Eqs.~(\ref{p1perppeak}), (\ref{p1longpeak}); (blue triangles) the exact numerical calculations. (a), (b) the  longitudinal R-CMT and (b), (d)  the transverse R-CMT. (c), (d) the relative errors with respect to the numerical values. The parameters are $I_p =0.5$ a.u., $p_{\bot i}=0.2$ a.u. and used cosinusoidal field had $E_0= 0.041$ a.u., $\omega = 0.0134$ a.u.}
    \label{fig:02:peak_comparison_corrections-E_0_04}
\end{figure}
\begin{figure}
	\centering
  		  		\includegraphics[width=1\linewidth]{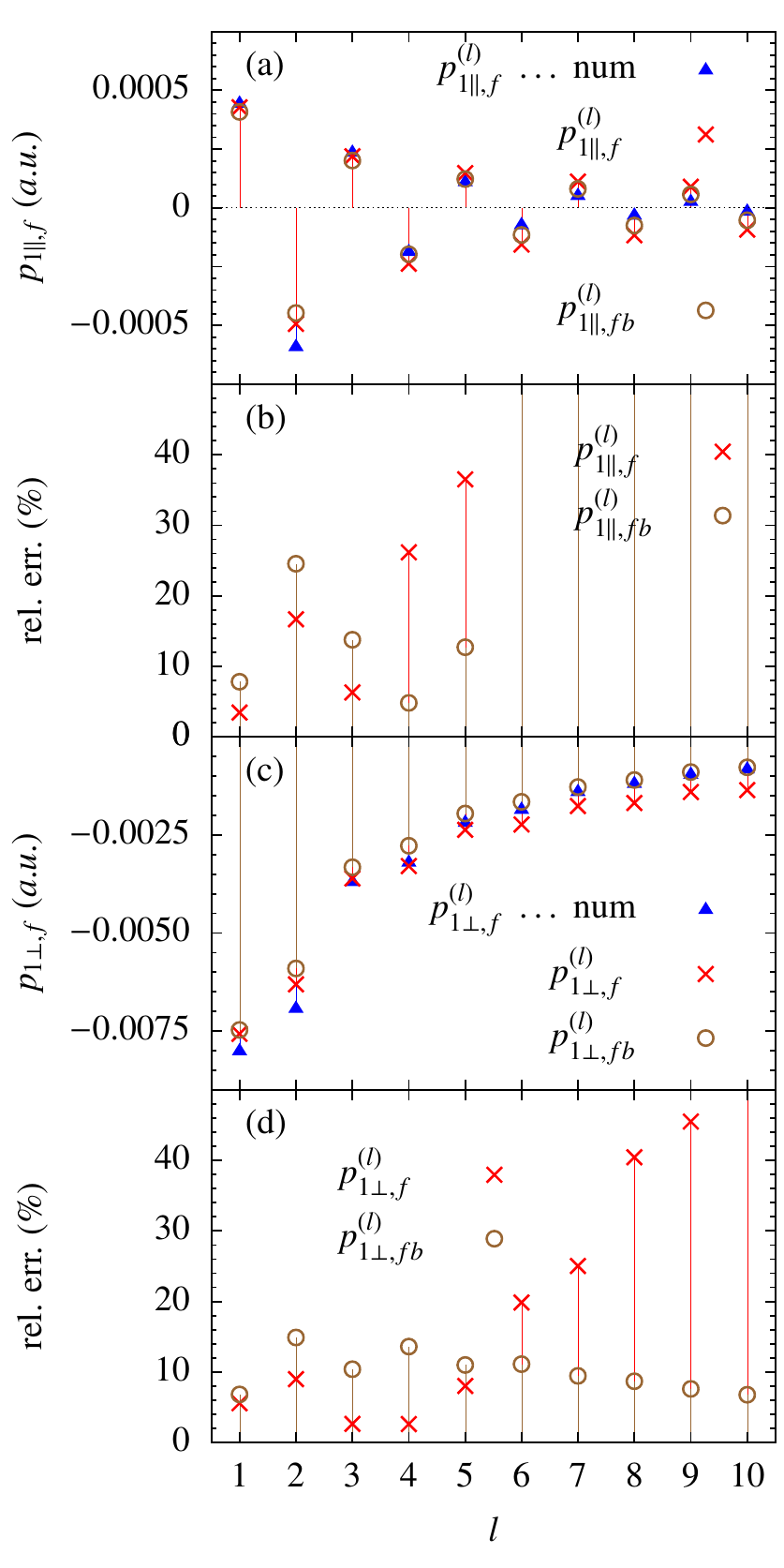}
  		    \caption{Graphical analysis of accuracy of R-CMT estimations at the $l^{\text{th}}$ FR: (brown circles) estimations via  Eqs.~(\ref{eq:12_px_corr_peak_preciser_turnpoints}), (\ref{eq:12_pz_corr_peak_preciser_turnpoints});  (red crosses) estimations via (\ref{eq:11:px_corr_plateau_preciser}), (\ref{eq:11:pz_corr_plateau_preciser}); (blue triangles) the exact numerical calculations. (a), (b)  the longitudinal momentum and (c), (d)  the  transverse momentum. (b), (d) the relative errors with respect to the numerical values. The parameters are the same as in Fig.~\ref{fig:02:peak_comparison_corrections-E_0_04}. }
    \label{fig:03:plateau_comparison_corrections-E_0_04}
\end{figure}

Note that the breakdown of the recollision picture can be observed also for SR. In fact, in  Fig. \ref{fig:02:peak_comparison_corrections-E_0_04}  the accuracy of the longitudinal momentum transfer estimation starts to increase at $11^{\text{th}}$ SR. However, the deviation from the exact numerical calculation is not as large as in the FR case. The reason is that the decay of the argument under the integrals in Eq. (\ref{eq:01:03:first_order:momentum_trans}) is much weaker in the case of FR than in the case of SR, which is due to the different orders of the leading terms in Eqs. (\ref{eq:SR_leading_order_x_z}) and (\ref{eq:FR_leading_order_x_z}).  

\subsection{I-CMT}\label{App:accuracy-I-CMT}

We compare our results for I-CMT with numerical calculation results in Figs.~\ref{fig:init_mom_corr_0_05} and \ref{fig:init_mom_corr_0_2} for $p_{\perp i} = 0.05$ and $p_{\perp i} = 0.2$, respectively. As we can see the next-to-leading order corrections to I-CMT significantly decrease the error of the estimation for both cases.
The corrected formulas manifest relative errors less than 5\% near the peak of the laser field, where the I-CMT effect is most significant.

\begin{figure}
		\begin{center}
	\includegraphics[width=0.98\linewidth]{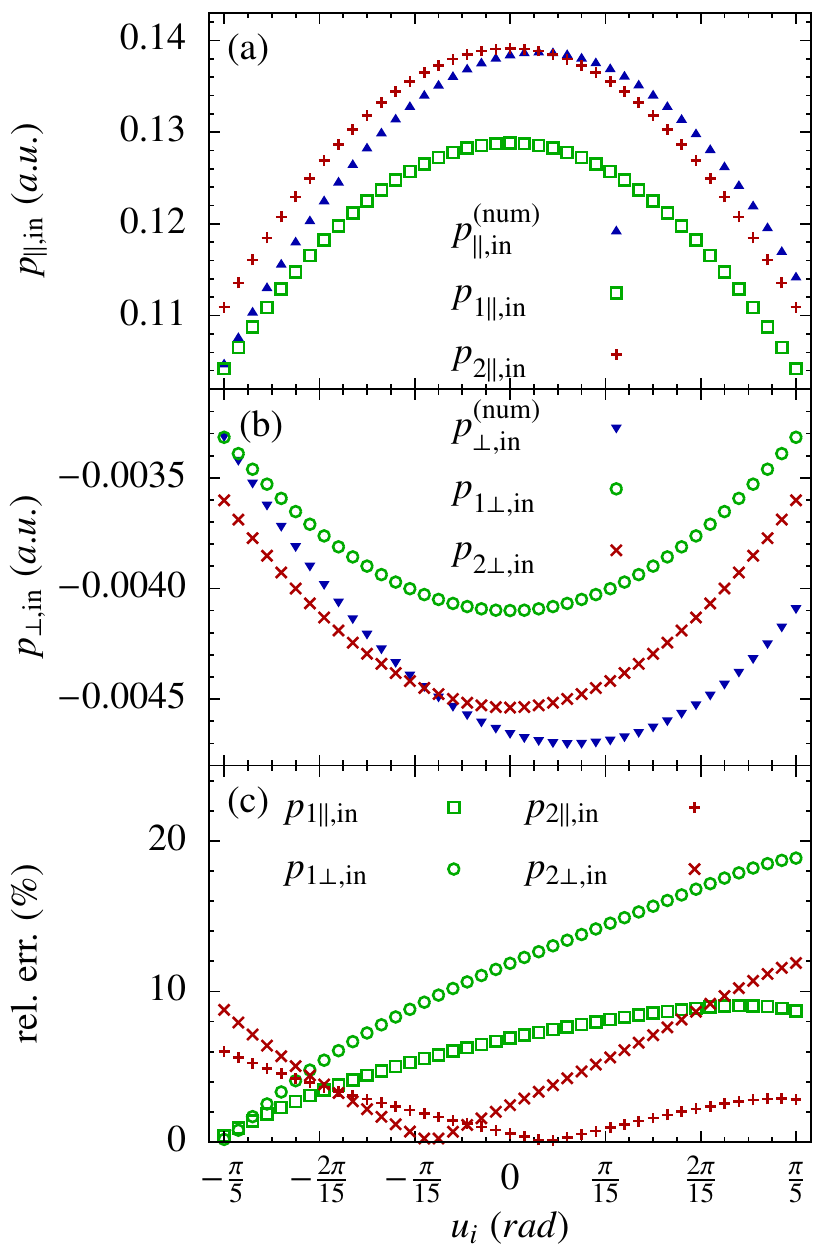}
	\end{center}
\caption{Estimation of I-CMT vs ionization phase: (a) transverse I-CMT; (b) longitudinal I-CMT; (c) relative error with respect to the exact numerical simulations,  for  the transverse I-CMT (red), and for the the longitudinal I-CMT (blue).  In (a),(b) blue triangles correspond to numerical simulations, green squares and circles to first-order I-CMT, and red  pluses and crosses to the corrected I-CMT. The parameter used are $p_{\perp i} = 0.05$, $E_0 = 0.041$, and $\omega = 0.0134$.}
\label{fig:init_mom_corr_0_05}
\end{figure}
\begin{figure}
		\includegraphics[width=0.98\linewidth]{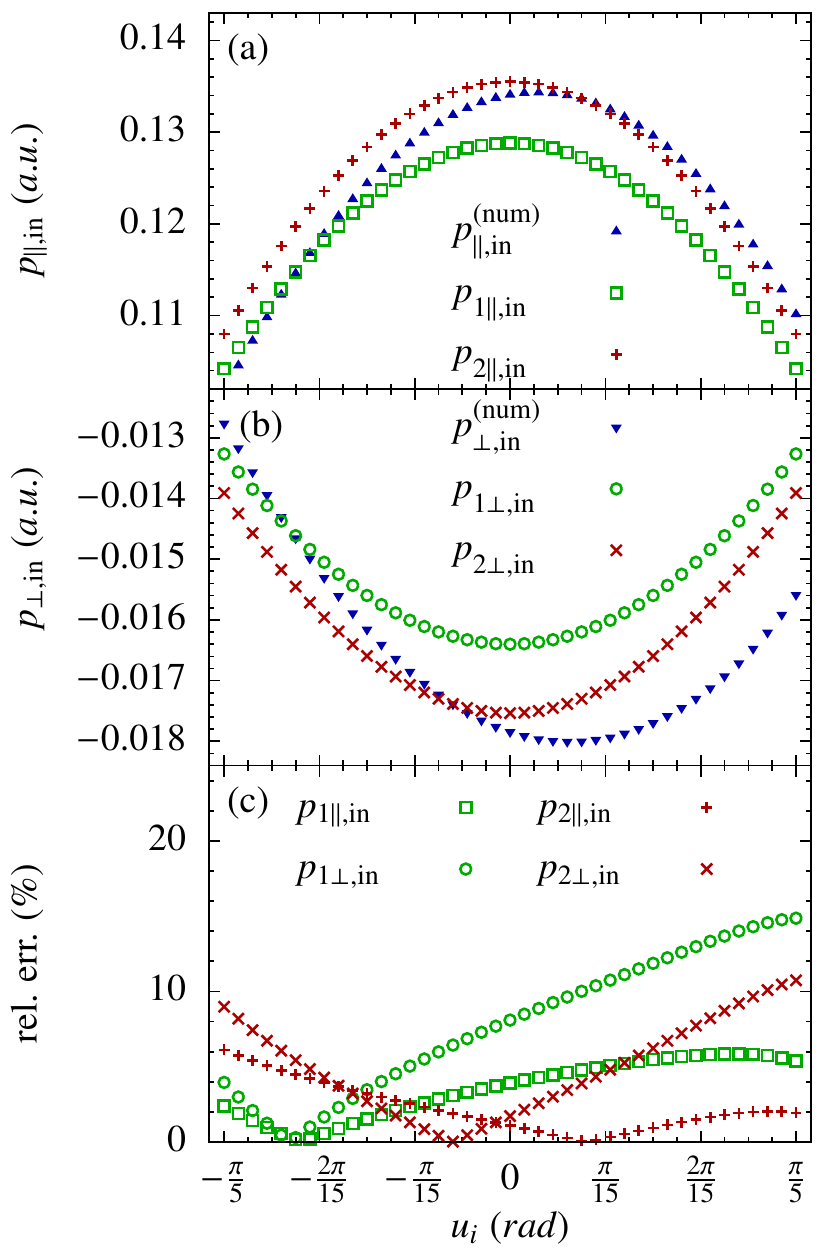}
\caption{Estimation of I-CMT vs ionization phase: (a) transverse I-CMT; (b) longitudinal I-CMT; (c) relative error with respect to the exact numerical simulations,  for  the transverse I-CMT (red), and for the the longitudinal I-CMT (blue). In (a), (b) blue triangles correspond to numerical simulations, green squares and circles to first-order I-CMT, and red plusses and crosses to the corrected I-CMT.  The parameter used are $p_{\perp i} = 0.2$, $E_0 = 0.041$, and $\omega = 0.0134$.}
\label{fig:init_mom_corr_0_2}
\end{figure}

\section{Methods}
\label{App:methods}
Despite the precise method we use, the essence of our semi-analytical approach is the correct determination of the rescattering points. The main challenge is to find an algorithm that will correctly choose the rescattering points and refuse other candidates. Even though the underlying principles for the rescattering point classification are the same, differences can be found and both methods are requiring special algorithms. Let us therefore discuss the methods separately. 

\subsection{Zero-order trajectory}
\label{App:methods_zero}
Since we know the whole zero-order trajectory distorted by the I-CMT analytically from Eqs. (\ref{eq:13:coordinate_init_distorted_x})-(\ref{eq:13:coordinate_init_distorted_z}), we can easily determine the rescattering points as $p_{0\parallel}(u_r) + p_{1\parallel,\text{in}} = 0$ for slow rescattering (save them in sorted list S) and $\tilde{x}_{0}(u_r) = 0$ for fast cases (and save them in sorted list F) up to the end of the laser pulse. The lists are sorted increasingly with respect to the rescattering phase $u_r$ of the individual events. The number of elements in a list will be denoted by prefix ``$\#$''. Once we have gathered all the candidates, we have to apply selection rules:
\begin{enumerate}
	\item remove the first element in S if its rescattering phase $u_r \sim \pi$
	\item when $\# \text{F} > 1$: starting from the lowest $u_r$, for each neighboring pair of F points find the S recollision they surround if its distance $|x(u_r)| > x_{\text{thresh}}$, otherwise memorize the two F points for later removal; finally, after analyzing of all F pairs remove all memorized F points
	\item from the S rescatterings we keep only those fulfilling one of the following two conditions:
	\begin{eqnarray}
	\mathbf{r}(u_r)\cdot\mathbf{E}(u_r) > 0 & \;\;\text{ and }\;\; & |x(u_r)| < x_{\text{thresh}}, \\
	\mathbf{r}(u_r)\cdot\mathbf{E}(u_r) < 0 & \;\;\text{ and }\;\; &|\mathbf{E}(u_r)| > E_{\text{thresh}},
	\end{eqnarray}
	where the first condition selects all the unfavorable turning points when they happen close enough to the ion and the second condition selects all the favorable turning points excluding those at the end of the laser pulse with respect to the thresh $E_{\text{thresh}}$.
\end{enumerate}
Let us note that the end of the pulse must be chosen approximately because otherwise fast recollisions could appear for late phases when the quiver motion is nearly non-existing. For such recollisions we lack estimations of R-CMT but their contribution to the total momentum is assumed to be vanishing due to the large distance from the ion for trajectories of our interest.

\subsection{Step-by-step}
\label{App:methods_step}
The step-by-step method is a little bit more elaborated and the general scheme from the zero-order trajectory cannot be used. Since we are interested only in the next rescattering point, there is only little sense to search for all the rescattering points till the end of the laser pulse in order to find the propper subsequent one. 

Therefore, we find all the slow or fast candidates within the range $u_r < 9\pi/4 + u_{in}$ with $u_{in} = u_{i}$ and $n=0$ by using the zero-order vector Eq.~(\ref{13:Step:iterative_p}) and the step-by-step evolved trajectory by Eqs.~(\ref{13:Step:iterative_x})-(\ref{13:Step:iterative_z}) in complete analogy to the zero-order case, prepend the ionization phase to the list F and apply the following procedure:
\begin{enumerate}
	\item if $\#\text{S} + \# \text{F} = 1$ there are no further rescattering points
	\item if the first event in F precedes the first event in S: remove the first element in S if its $u_r < \pi +  u_{in}$
	\item if the first element in S precedes the first element in F: 
	\begin{enumerate}
		\item remove the second element in S if its $u_r < \pi/4 + u_{in}$ (eliminating fake slow recollision due to non-negligible R-CMT)
		\item if $\# \text{F} > 0$ then remove the first element in F when its $u_r < \pi + u_{in}$, otherwise remove every even element in S or every element in S for which $|\mathbf{E}(u_r)| < E_{\text{thresh}}$; the remaining list S contains all valid rescattering points
	\end{enumerate}
	\item when $\# \text{F} > 1$: starting from the lowest $u_r$, for each neighboring pair of F points find the S recollision they surround if its distance $|x(u_r)| > x_{\text{thresh}}$, otherwise memorize the two F points for later removal; finally, after analyzing of all F pairs remove all memorized F points
	\item from the S rescatterings we keep only those fulfilling one of the following two conditions:
	\begin{eqnarray}
	\mathbf{r}(u_r)\cdot\mathbf{E}(u_r) > 0 & \;\;\text{ and }\;\; & |x(u_r)| < x_{\text{thresh}}, \\
	\mathbf{r}(u_r)\cdot\mathbf{E}(u_r) < 0 & \;\;\text{ and }\;\; &|\mathbf{E}(u_r)| > E_{\text{thresh}},
	\end{eqnarray}
	where the first condition selects all the unfavorable turning points when they happen close enough to the ion and the second condition selects all the favorable turning points excluding those at the end of the laser pulse with respect to the thresh $E_{\text{thresh}}$.
\end{enumerate}
At the end of this procedure, we have a list of all valid rescattering points including the original one. The second element of the union S + F is the next rescattering point and we can estimate the momentum after this rescattering from Eqs. (\ref{13:Step:iterative_p}) by using the right rescattering type and formula. Now we set $u_{in} = u_r$ and clear the lists S, F; further, we append the rescattering point $u_r$ to the proper list instead of the ionization phase and repeat the whole procedure above for $n = n + 1$. We repeat so till no further rescattering points are found (i.e., $\# S + \# F = 1$) or until the end of the laser pulse. Again, there is some caution required while setting the end of the pulse in order to eliminate the unwanted fast recollisions.

\bibliography{strong_fields_bibliography.bib}

\end{document}